\documentclass[twocolumn,preprintnumbers,amsmath,amssymb,floatfix]{revtex4}

\usepackage{graphicx}
\usepackage{dcolumn}
\usepackage{bm}
\usepackage{graphics}
\usepackage[usenames]{color}
\usepackage{hyperref}

\hypersetup{
    bookmarks=true,         
    unicode=false,          
    pdftoolbar=true,        
    pdfmenubar=true,        
    pdffitwindow=false,     
    pdfstartview={FitH},    
    pdftitle={My title},    
    pdfauthor={Author},     
    pdfsubject={Subject},   
    pdfcreator={Creator},   
    pdfproducer={Producer}, 
    pdfkeywords={keywords}, 
    pdfnewwindow=true,      
    colorlinks=true,       
    linkcolor=red,          
    citecolor=blue,        
    filecolor=magenta,      
    urlcolor=green           
}
\newcommand {\apgt} {\ {\raise-.5ex\hbox{$\buildrel>\over\sim$}}\ }
\newcommand {\aplt} {\ {\raise-.5ex\hbox{$\buildrel<\over\sim$}}\ }
\begin{document}

\preprint{}

\title{Characterization of the Dynamics of Glass-forming Liquids from the Properties of the Potential Energy Landscape}

\author{ Sumilan Banerjee, Chandan Dasgupta}
 \affiliation{Department of Physics, Indian Institute of Science, Bangalore}

\begin{abstract}
We develop a framework for understanding the difference between \emph{strong} and
\emph{fragile} behavior in the dynamics of glass-forming liquids from the properties 
of the potential energy landscape.
Our approach is based on a master equation 
description of the activated jump dynamics among the local minima of the potential energy 
(the so-called \emph{inherent structures})
that characterize the potential energy landscape of the system. 
We study the dynamics of a small
atomic cluster using this description as well as molecular dynamics simulations and 
demonstrate the usefulness of our
approach for this system. Many of the remarkable features of the complex dynamics of glassy
systems emerge from the activated dynamics in the potential energy landscape of the atomic cluster. 
The dynamics of the system  exhibits typical characteristics of a
\emph{strong} supercooled liquid when the system is allowed to explore
the full configuration space. This behavior arises because the dynamics is dominated by a few lowest-lying
minima of the potential energy and the potential energy barriers between these minima. 
When the system is constrained to explore only a limited region of the potential energy landscape that excludes
the basins of attraction of a few lowest-lying minima, the dynamics is found to exhibit the characteristics of
a \emph{fragile} liquid.
\end{abstract}
\maketitle

\section{Introduction}

In the supercooled state, glass-forming liquids exhibit many fascinating features 
\cite{CAAngell1,WKob,PGDebenedetti} in its dynamic behavior, such 
as multistage, non-exponential decay of fluctuations and a rapid growth of relaxation times
with decreasing temperature. In
these aspects, glassy systems challenge us with many interesting issues and questions which are not well 
resolved theoretically.

 A popular phenomenological characterization of the dynamics of glassy systems, proposed by Angell,
\cite{CAAngell1,WKob,PGDebenedetti,CAAngell2} is the classification of the dynamics as \emph{strong} or \emph{fragile}. 
It classifies different glass
formers on the basis of the temperature dependence of their viscosity $\eta$ or their 
structural relaxation time $\tau$. Quite generally, the rapid growth of $\tau$ with temperature $T$ can be
represented by a generalized Arrhenius form, $\tau(T)\sim \exp{(E_b(T)/k_BT)}$ with an effective temperature
dependent activation energy $E_b(T)$. For \emph{strong} systems, this activation energy is essentially independent of
temperature while \emph{fragile} systems exhibit a strong temperature dependence of this quantity: it increases as $T$
is decreased. This implies that
the relaxation mechanism is independent of temperature for the first type of systems, whereas for the other
type, it depends on $T$. Although various measures of the extent of \emph{fragility} exist in literature
\cite{GRuocco1} in terms of the slope of the $\ln(\tau)$ vs. $1/T$ plot, the quantitative distinction between strong and
fragile liquids at a microscopic level is not fully understood yet. 

Following Goldstein \cite{MGoldstein1}, a widely accepted way of looking at glassy dynamics
is to view the dynamical evolution of the system in terms of the motion of a state point in configuration space, specified by
$3N$ coordinates for an $N$ particles system, over its potential energy surface, often  referred to
as potential energy landscape (PEL) \cite{DJWales1}. For glass-forming liquids and other disordered systems such as spin 
glasses \cite{KBinder} in general, a generic feature of the PEL is the existence of a large
number of local minima in it, so that the potential energy surface has very \emph{rough} topography. At low temperature, in the 
supercooled regime, the system visits the neighborhood of a 
local minimum for very long times and makes occasional jumps to other minima close to the initial one over the barriers
separating them. Following this description, Stillinger and Weber \cite{FHStillinger1,FHStillinger2,FHStillinger3} showed that a useful
approach towards the understanding of the low-temperature properties of such system with a rugged PEL is to divide the configuration space into 
basins of attraction of the local minima and then formulate a statistical description in terms of the distributions of 
different properties of these local
minima, denoted as \emph{inherent structures} (IS), and their basins of attraction. 
Stillinger also suggested \cite{FHStillinger4} a rationale for the difference
between strong and fragile liquids based on the PEL viewpoint. According to his idea, strong liquids have
a uniformly rough PEL and there is not much variation in the values of the energy barriers separating different inherent structures. 
On the contrary, the PEL of
fragile liquids have non-uniform roughness. At high temperatures, the system explores a PEL with nearly uniform roughness due to 
its high kinetic energy, but at lower temperatures it explores the deep valleys with very different energy barriers,
giving rise to a temperature-dependent average barrier height. The work of Sastry et al.~\cite{SSastry1}
demonstrated the usefulness of this description, although the understanding of the dynamics in this approach
still remains qualitative to a large extent. 

After early work \cite{FHStillinger1,FHStillinger2} on the implementation of 
a description based on inherent structures by supplementing conventional Molecular Dynamics (MD) simulations
\cite{DFrenkel} with regular 
steepest-descent minimizations of the potential energy (called \emph{quenches}) and thereby sampling the inherent structures, 
a large amount of activity has gone into the field in recent years and a variety of methods have been suggested to make the survey 
of the PEL more efficient \cite{DJWales1}. Once the inherent structures are sampled properly, rates of transition between 
them may also be calculated and the dynamics of the system can be described by the evolution of the probabilities of
occupying different basins through a master equation \cite{DJWales1,LAngelani1,LAngelani2,MAMiller1}. An important assumption 
behind such an approach is a separation of time scales i.e.~the assumption that the intra-basin relaxation time $\tau _\mathrm{intra}$ 
and the inter-basin 
relaxation time $\tau _\mathrm{inter}$ are well separated, $\tau_\mathrm{intra} \ll \tau_\mathrm{inter}$. This is indeed
a good assumption at low temperatures for which the typical barrier height $E_b>k_BT$. This formulation does not
have the usual limitations arising from the requirement of long simulation times, since the master equation can be
formally solved for all times. However, for writing down the rates of transition between two minima, we need
to sample the barriers or transition states between them and finding transition states is much more demanding
computationally than finding the local minima. However, in recent years, various methods \cite{DJWales1} have
been developed to overcome this difficulty. 

The number of minima, $n_{min}$ for model glass formers
increases exponentially with the number of particles in the system \cite{FHStillinger4}. So it is
impractical to hope to list all the minima even for a system of moderate size, say one consisting of a few hundred
particles. Moreover, the formal solution of the master equation requires the diagonalization of matrix of order
$n_{min}\times n_{min}$. Hence, the applicability of this method has been restricted mostly to small system sizes till
now \cite{DJWales1}. Alternatively one can organize a \emph{set} of inherent structures into larger
\emph{metabasin} \cite{BDoliwa1,AHeuer,YYang} and set up a master equation dynamics for transitions between the metabasin
\cite{YYang}. Another aim of the metabasin construction is to define a space where the dynamics of the
system point become Markovian, a crucial assumption behind any master equation based description. The
Markovian assumption might break down for elementary jumps between the basins of inherent structure. However
assigning the rate of transition between two metabasins is not straightforward as one needs to integrate 
out the intra-metabasin dynamics (e.g. jumps between basins in the same metabasin) in that case. Also, the construction of the metabasins
out of inherent structures is generally done using somewhat ad-hoc criteria \cite{BDoliwa1,AHeuer,YYang}. 

Nevertheless, the long-time, low-temperature dynamics of systems with a small number of particles 
(say ten to hundred) interacting via some model potential has been studied very efficiently
using this coarse-grained (in time) master equation approach since one can make almost exhaustive 
search of
all the minima (a few hundreds to several thousands) and obtain a moderately good number of transition states for
them \cite{DJWales1,TFMiddleton}. Though small, these clusters captures many features of the complex dynamics
\cite{MAMiller1} observed in larger systems and hence constitute a good playground for relating the properties of 
the PEL to the dynamics.

In the same spirit, we consider here the master equation dynamics in a connected \emph{network}
of minima \cite{LAngelani1}, where the minima serve as the \emph{nodes} and transition states as the \emph{edges} of the network
\cite{JPKDoye1}. In Section \ref{sec.MasterEquation}, we briefly review the general formalism \cite{LAngelani1,MAMiller1} 
for calculating time autocorrelation functions of various
physical quantities and the corresponding relaxation times based on the master equation dynamics in the network of
minima. We show that a quantitative understanding of \emph{fragility} can be obtained in
this framework from an analysis of elementary jumps between inherent structures and the effective barrier $E_b(T)$ appearing in
the temperature dependence of the relaxation time can be directly calculated from the local properties of the
minima and the transition states that connect them. We also comment on the breakdown of Stokes-Einstein
\cite{DJWales1} relation observed in many glass formers from this perspective.

To test the validity of our results, we study of the equilibrium dynamics of a cluster of 13 atoms interacting by the Morse 
potential \cite{PMMorse} in Sections \ref{sec.NetworkMorseCluster} and \ref{sec.DynamicsMorseCluster}. The PEL of
this system has been studied in detail in the past \cite{MAMiller2} and a nearly 
exhaustive list of minima and transition states has been obtained. The PEL resembles a funnel, in which the minima 
are organized into pathways of decreasing energy leading to the global minimum. This system has the nice property of 
having a complex landscape that consists of a fairly large but still manageable number of minima and also displaying 
some of the salient features of the complicated dynamics of glassy system, as we report in Section \ref{sec.DynamicsMorseCluster}. By restricting the 
system to sample certain parts of the PEL excluding a few lowest-lying minima, we are able to show that the dynamics of 
the system in this restricted part of the configuration space exhibits characteristic behavior of {\it fragile} systems,
i.e.~$E_b(T)$ is perceptibly temperature dependent, whereas the dynamics in the full PEL exhibits {\it strong} features. We have
also carried out MD simulation for the 13-atom Morse cluster. Using simulations in which the MD trajectories are confined
\cite{SFChekmarev1,SFChekmarev2}
in appropriately restricted parts of the PEL, we are able to substantiate the conclusions of the network model calculation.
The results from MD simulations are discussed and compared with those obtained from the network model in
Section \ref{sec.MD}. Some of the technical details of the network model calculations and restricted MD simulations 
are described in the Appendices
\ref{app.ActivationEnergy}, \ref{app.BuildNetwork} and \ref{app.IntervalBisection}.

\section{Master equation for jump dynamics between inherent structures} \label{sec.MasterEquation}

 The first step in exploring the landscape is to find the configuration corresponding to the local minima and the
transition states. These are the stationary points or \emph{saddles} of the potential energy
function $V(\mathbf{r}_1,\mathbf{r}_2,...,\mathbf{r}_N)$, characterized by $\nabla V=0$ and number of negative
eigenvalues of the Hessian matrix $\mathbf{H}\doteq V_{ij}^{\alpha\beta}=\frac{\partial^2 V}{\partial
r_i^\alpha \partial r_j^\beta}$, $r_i^\alpha$ being the $\alpha$-th coordinate of the $i$-th particle.
The eigenvector corresponding to each negative eigenvalue of the Hessian matrix at a stationary point signifies an unstable
direction and stationary points can be indexed by the number of such unstable directions. For instance, at a
local minimum, there is no negative eigenvalue and it can be denoted as saddle of index 0, a transition state as
saddle of index 1 or first-order saddle and similarly there can be higher order saddle points having index
running from 2 to the dimension of the configuration space. One
can use steepest descent minimization \cite{WHPress} for finding the minima and transition states can be located
efficiently using the eigenvector following method \cite{CJCerjan,Optim}. Once the minima and transition states that
connect them are known, we can arrange them by designating as \emph{nodes} and \emph{edges}, respectively, of a network. We
describe this procedure in greater detail later for the specific case of a 13-atom Morse cluster (see also
Appendix \ref{app.BuildNetwork}).

In the model of a connected network of potential energy minima, the master equation for the jump dynamics can
be written as
\begin{eqnarray}
\frac{dP_a(t;b,t_0)}{dt}&=&\sum_c W_{ac}P_c(t;b,t_0). \label{eq.MasterEquation}
\end{eqnarray}  
$P_a(t;b,t_0)$ is the probability that the system is at minimum $a$ at time $t$, if it was at a minimum $b$ at
time $t_0$ and $a$ runs from $1$ to $n_\mathrm{min}$, the total number of minima in the network. The
off-diagonal elements of the matrix $\mathbf{W}$ are the transition rates and, as usual, the diagonal elements
are fixed by the condition $\sum_a P_a(t;b,t_0)=1$ implying $\sum_a W_{ac}=0$. In order to obtain an
asymptotic behavior (in the long time limit) that agrees with the Boltzmann distribution, the occupation
probability must satisfy $\mathrm{lim}_{t\rightarrow \infty} P_a(t;b;t_0)=P^0_a\equiv
\mathcal{Z}^{-1}(\mathrm{Det}(\mathbf{H}_a))^{-1/2}\exp{(-V_a/T)}$. Here, $\mathcal{Z}$ is such that $\sum_a
P_a^0=1$ and the pre-exponential factor follows from a harmonic approximation for the partition function in the basin 
of each minimum (we take the
Boltzmann constant $k_B=1$). The matrix $\mathbf{H}_a$ is the Hessian matrix for the $a$-th minimum. As
$\mathbf{W}$ ought to satisfy the detailed balance relation $W_{ab}P_b^0=W_{ba}P_a^0$, in numerical
calculation it is more convenient to express the solution in terms of the eigenvectors of the real symmetric
matrix $\widetilde{\mathbf{W}}_{ab}\equiv \mathbf{W}_{ab}(P_b^0/P_a^0)^{1/2}$. Finally, one can formally obtain
the solution of the master Equation \eqref{eq.MasterEquation} for all time,
i.e.~$P_a(t;b,t_0)=(P_a^0/P_b^0)^{1/2}\sum_n e_a^{(n)}e_b^{(n)}e^{\lambda_n(t-t_0)}$. Here
$\mathbf{e}^{(n)}$ are the eigenvectros of $\widetilde{\mathbf{W}}$ corresponding to the eigenvalue
$\lambda_n$ ($1\leq n\leq n_\mathrm{min}$). The matrix $\widetilde{\mathbf{W}}$ has one zero eigenvalue
corresponding to the equilibrium distribution and all other eigenvalues are negative. We shall follow the
convention of arranging $\lambda_n$'s in descending order (ascending order in their absolute values) starting
from $\lambda_1=0$.

The model is well-defined once we give an appropriate expression for the transition or \emph{hopping} matrix
$\widetilde{\mathbf{W}}$ between the nodes of the network of minima, Treating the problem as a  Markovian
Brownian multi-dimensional motion in the over-damped friction regime \cite{HAKramers,JSLanger,PHanggi}, we can write the transition rates
between a directly connected or nearest-neighbor pair of minima, $<ab>$, from $b$ to $a$ over the saddle $s$
as
\begin{eqnarray}
W_{ab}^s&=&
\frac{\tilde{\omega}_{s,ab}^2}{\mu}\sqrt{\frac{\mathrm{Det}(\mathbf{H}_a)}{|\mathrm{Det}(\mathbf{H}_{ab}^s)|}}e^{-\frac{V_{ab}^s-V_b}{T}}.
\label{eq.TransitionMatrix}
\end{eqnarray}
Here, $\tilde{\omega}_{s,ab}$ is the down frequency at the saddle point i.e.~$\tilde{\omega}_{s,ab}^2=\Lambda_{ab}^s$,
$\Lambda_{ab}^s$ being the magnitude of negative eigenvalue of the Hessian matrix $\mathbf{H}^s_{ab}$ at the transition
state, $\mu$ is the friction constant that sets the time scale (we take $\mu=1$ in all our calculation
henceforth), and $V_b$, $V_{ab}^s$ are, respectively, the potential energies at the minimum $b$ and the saddle point $s$ between
minima $a$ and $b$. If there are multiple barriers connecting $b$ and $a$ then the total transition rate from $b$ to
$a$, $W_{ab}$ is obtained by summing over all the barriers i.e.~$W_{ab}=\sum_{s\in <ab>} W_{ab}^s$.

\subsection{Correlation function} \label{subsec.CorrelationFunction}

 To study the dynamics and calculate relevant relaxation times in the network model, one needs to define the equilibrium time
autocorrelation function of some physical quantity, say $\phi(\mathbf{r}(t))$, a generic observable which
depends on the collective coordinate $\mathbf{r}(t)=\{\mathbf{r}_1(t),\mathbf{r}_2(t),...,\mathbf{r}_N(t)\}$
at times $t$. In this language, the time autocorrelation function can be written \cite{LAngelani1,LAngelani2} as
\begin{eqnarray}
&&\langle\phi(\mathbf{r}(t))\phi(\mathbf{r}(t_0))\rangle=\langle\Phi(t,t_0)\rangle=C_\phi(t,t_0) \nonumber\\
&&=\sum_b P^0_b\sum_a \Phi_{ab}P_a(t;b,t_0)\nonumber\\
&&=C_\phi^0+\sum_{n\geq 2}e^{\lambda_n(t-t_0)}\sum_{a,b}\Phi_{ab}(P^0_aP^0_b)^{1/2}e_a^{(n)}e_b^{(n)}.\label{eq.CorrelationFunction}
\end{eqnarray}  
Here $\Phi_{ab}=\Phi(\mathbf{r}_a,\mathbf{r}_b)=\phi(\mathbf{r}_a)\phi(\mathbf{r}_b)$ and
$C_\phi^0=\sum_{a,b}\Phi_{ab}P^0_aP^0_b$. Also, $C_\phi(t=t_0,t_0)=\sum_a \Phi_{aa}P^0_a$ and
$\lim_{t\rightarrow\infty}C_\phi(t,t_0)=C_\phi^0$ are short-time and long-time limits of the correlation
function $C_\phi(t,t_0)$, respectively. 

Once the correlation function is calculated, the relaxation time can be estimated by assuming a pure Debye
(single exponential) relaxation, such that $\tilde{C}_\phi(t-t_0)=C_\phi(t,t_0)-C_\phi^0\equiv
\tilde{C}_\phi(t_0)\exp{(-(t-t_0)/\tau_\phi^e)}$ and evaluating the area under the resulting curve from
Eq.\eqref{eq.CorrelationFunction}, i.e.
\begin{eqnarray}
\tau_\phi^e=-\frac{\sum_{n\geq2}\frac{1}{\lambda_n}\sum_{a,b}(P_a^0P_b^0)^{1/2} \Phi_{ab}
e_a^{(n)}e_b^{(n)}}{\sum_{n\geq2,a,b}(P_a^0P_b^0)^{1/2} \Phi_{ab} e_a^{(n)}e_b^{(n)}}.
\end{eqnarray}
This way of defining the relaxation time relies on the assumption that the decay of $C_\Phi(t-t_0)$ is well
described by a single exponential, but in glassy systems the temporal decay of correlation often follows a
profile that is more complex than a simple exponential. The precise form of the decay is usually not known, although it can be
fitted with the empirical Kohlrausch-Williams-Watts (KWW) or stretched exponential function
\cite{RKohlrausch,GWilliams}
\begin{eqnarray}
\tilde{C}_\phi(t)=\tilde{C}_\phi^0e^{-\left(t/\tau_\phi\right)^\beta} \label{eq.KWW}
\end{eqnarray} 
in many cases. Here $\tau_\phi$ is a measure of the relaxation time and $\beta$ is the KWW (or stretching) exponent. The above form
[Eq.\eqref{eq.KWW}] and other typical features of glassy dynamics such as the decoupling of transport
coefficients (i.e.~the violation of the Stokes-Einstein relation $\eta\propto T/D$, $\eta$ and $D$ being the
viscosity and diffusion coefficient, respectively) can be rationalized from the hypothesis of the existence of
a distribution of time scales that is not sharply peaked at a particular value. A distribution of time scales may arise 
e.g.~from \emph{dynamical heterogeneity} that describes spatial variations of the local relaxational dynamics (i.e. the fact that
different parts of the sample may have different relaxation times). Given a distribution of relaxation times,
$\rho^\phi(\tau)$, the temporal decay of a typical time autocorrelation function, reflecting
the effects of all these relaxation processes, is given by 
\begin{eqnarray}
\tilde{C}_\phi(t)&=&\int_0^\infty
e^{-\left(t/\tau\right)}\rho^\phi(\tau)d\tau, \label{eq.DynamicalHeterogeneity}
\end{eqnarray}
and one can easily show that this results in a stretched
exponential form for an appropriate distribution $\rho^\phi(\tau)$.
In the present master equation framework, a distribution of relaxation time is naturally provided by the
different modes $n=2,...,n_\mathrm{min}$ and one can identify the contribution or \emph{weight} [similar to
$\rho^\Phi(\tau)$ in Eq.\eqref{eq.DynamicalHeterogeneity}] of the $n$-th mode as
\begin{eqnarray}
w_\phi^{(n)}=\frac{\sum_{a,b}(P_a^0P_b^0)^{1/2}\Phi_{ab}e_a^{(n)}e_b^{(n)}}{\sum_{n\geq2,a,b}(P_a^0P_b^0)^{1/2}\Phi_{ab}e_a^{(n)}e_b^{(n)}},\label{eq.ModeWeight}
\end{eqnarray}
where $\tau_n\equiv|\lambda_n|^{-1}$ is the relaxation time corresponding to the $n$-th mode. A distribution of
relaxation times in the jump dynamics on the PEL  appears due to a \emph{heterogeneous} distribution of
barrier heights  contributing to different relaxation \emph{channel} $n$.
Though the origin of non-exponential relaxation due to a broad distribution of relaxation times in the master
equation based framework looks superficially similar to that due to dynamical heterogeneity mentioned above,
the precise correspondence between the \emph{heterogeneity} of barrier heights in the configuration space and
\emph{dynamical heterogeneity} in the real space is not very clear.

As shown in Appendix \ref{app.ActivationEnergy}, one can utilize Eq.\eqref{eq.ModeWeight} to directly calculate the temperature dependent
activation energy $E_b(T)$ [or, more precisely $E_b^\phi(T)$] to a good approximation and a quantitative measure, namely
$\mathcal{W}_{ab}^\phi$, of the participation of individual barriers (or the pairs of minima that are connected by
the barrier) in the relaxation process. We have tested these results [Eqs. \eqref{eq.EffectiveBarrier} and
\eqref{eq.MinimaPairWeight} in Appendix
\ref{app.ActivationEnergy}] for the case of 13-atom Morse cluster in Section \ref{sec.DynamicsMorseCluster}.

\subsection{Structural relaxation} \label{subsec.Structuralrelaxation}

Since we are mainly interested in structural relaxation of the system as its state point explores different
regions of the PEL, we look for the time autocorrelation function of quantities related to the configurations
of the minima. One such quantity whose autocorrelation function and related relaxation time is of much
interest in glass physics is the off-diagonal microscopic stress tensor \cite{JPHansen},
i.e.~$\sigma^{\alpha\beta}\equiv \sum_i v_i^\alpha v_i^\beta-\sum_{i<j} V'(r_{ij})(r_{ij}^\alpha
r_{ij}^\beta/r_{ij})$ ($\alpha \neq \beta$). Here, $v_i^\alpha$ and $r_{ij}^\alpha$ are the $\alpha$-th
components of the velocity of the $i$-th particle and $\mathbf{r}_{ij} \equiv \mathbf{r}_i-\mathbf{r}_j$, respectively, while
$V'(r)=\frac{\partial V(r)}{\partial r}$. Since the velocity $\mathbf{v}_i$ is not defined in the model, following
reference \cite{LAngelani1}, we neglect the kinetic energy term and work with the following quantity
\begin{eqnarray}
\sigma^{\alpha\beta}&=&-\sum_{i<j}V'(r_{ij})\frac{r_{ij}^\alpha
r_{ij}^\beta}{r_{ij}}. \label{eq.StressTensor}
\end{eqnarray}
Consequently, we can calculate the stress-stress autocorrelation function
$C_\sigma(t)=(1/3)\sum_{\alpha<\beta}\langle\sigma^{\alpha\beta}(t)\sigma^{\alpha\beta}(0)\rangle$ from Eq.\eqref{eq.CorrelationFunction}
 where we need to insert $\Phi_{ab}=\sum_{\alpha<\beta}
\sigma_a^{\alpha\beta}\sigma_b^{\alpha\beta}$. The shear viscosity, $\eta\propto T^{-1}\int_0^\infty dt
C_\sigma(t)$, is related to the stress autocorrelation function. We compute another correlation function,
termed as \emph{overlap function} $C_\delta(t)$, by setting $\Phi_{ab}=\Phi(\mathbf{r}(t)\in \mathcal{B}_a,\mathbf{r}(0)\in
\mathcal{B}_b)=\delta_{ab}$ in Eq.\eqref{eq.CorrelationFunction} to meaningfully compare the prediction of the network model 
with MD (see Section \ref{sec.MD}). The basin of attraction $\mathcal{B}_a$ of the $a$-th minimum is the set of
state points that flow to the $a$-th minimum under steepest descent minimization and
$\delta_{ab}$ is the usual Kronecker delta function. The correlation function $C_\delta(t)$ decays solely due to
transitions between inherent structures and it can be calculated from MD simulation as we show in Section
\ref{sec.MD}.

\begin{figure*}
\begin{center}
\begin{tabular}{ccc}
\includegraphics[width=6cm]{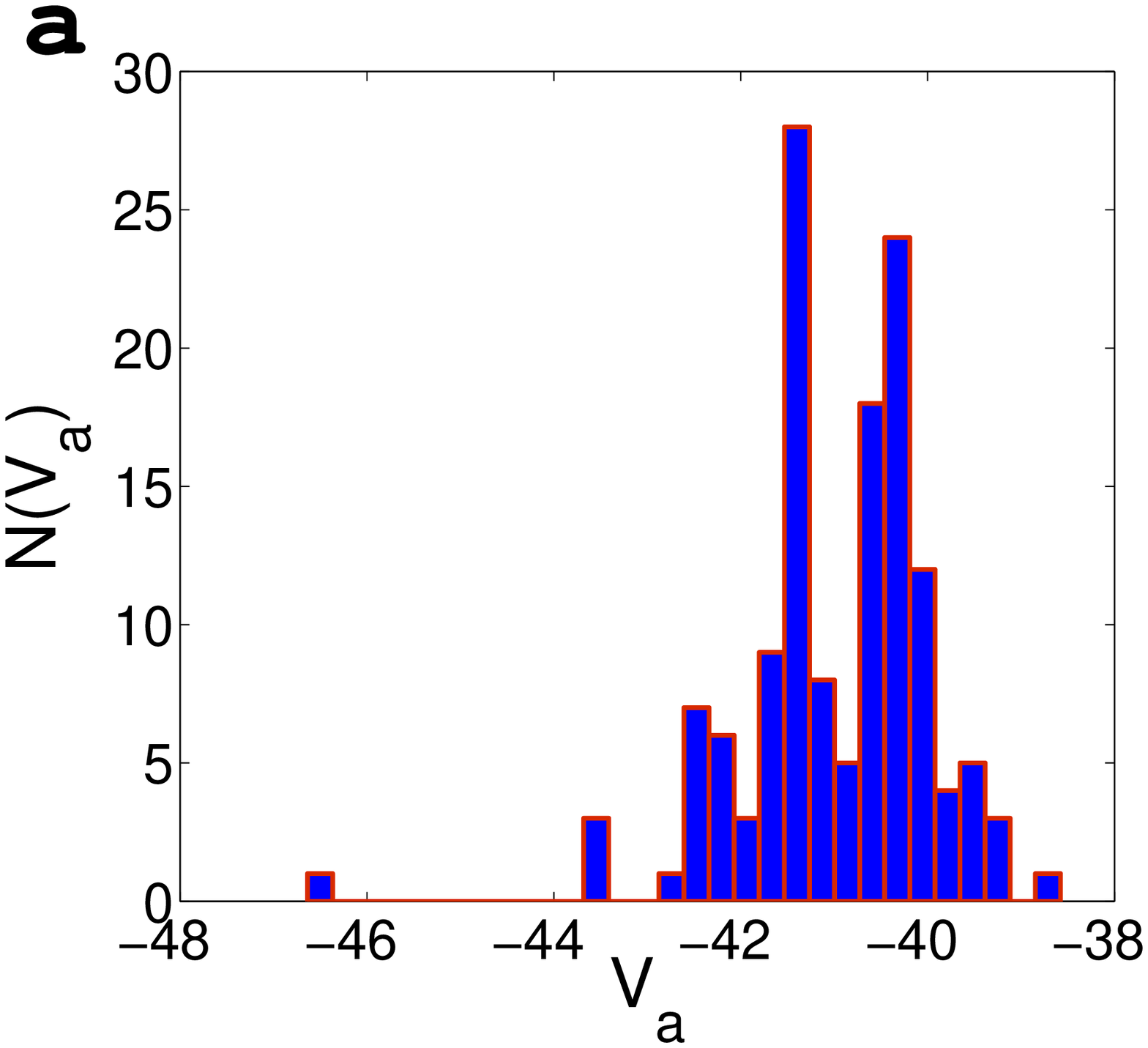}&
\includegraphics[width=6cm]{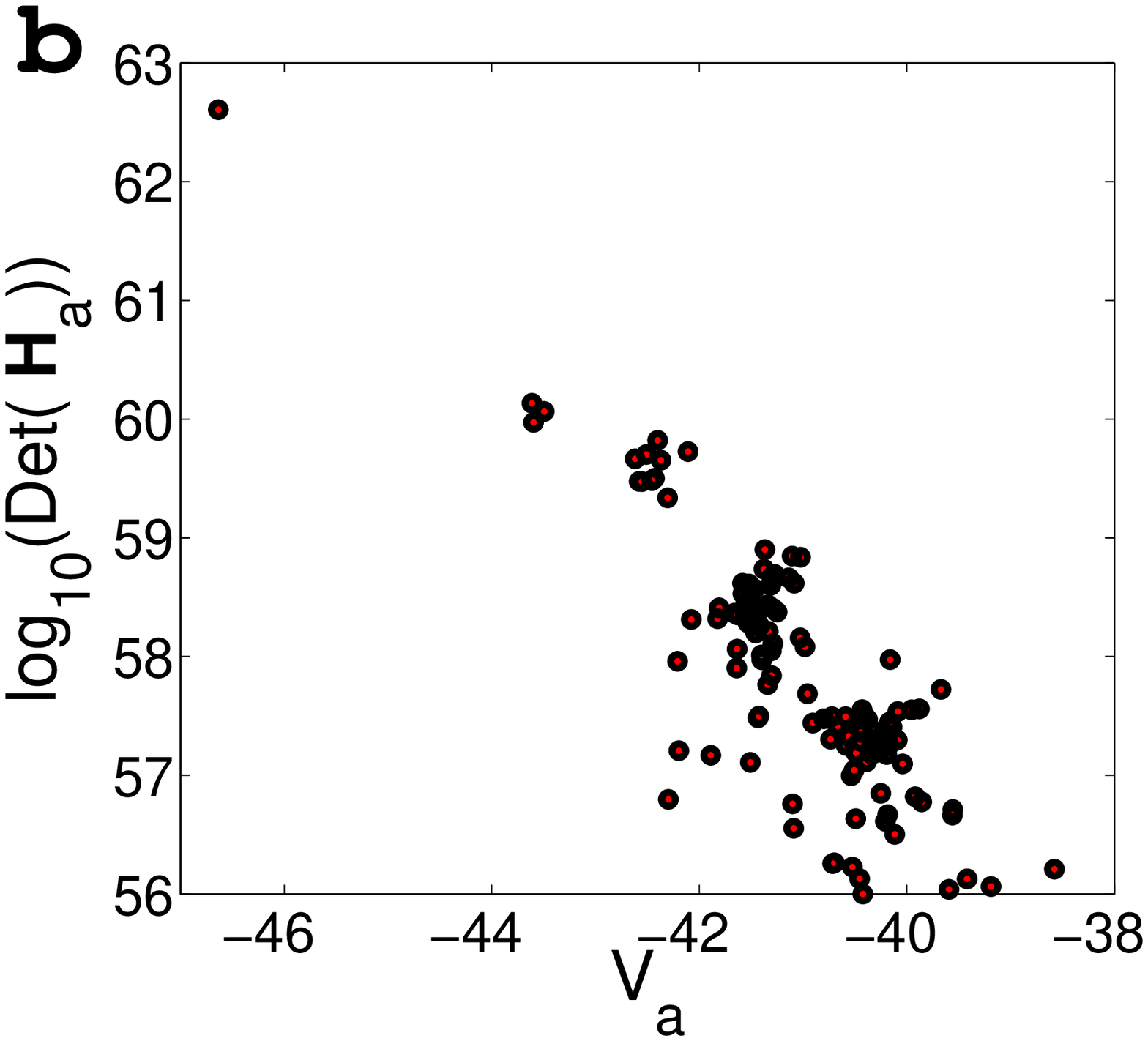}&
\includegraphics[width=6cm]{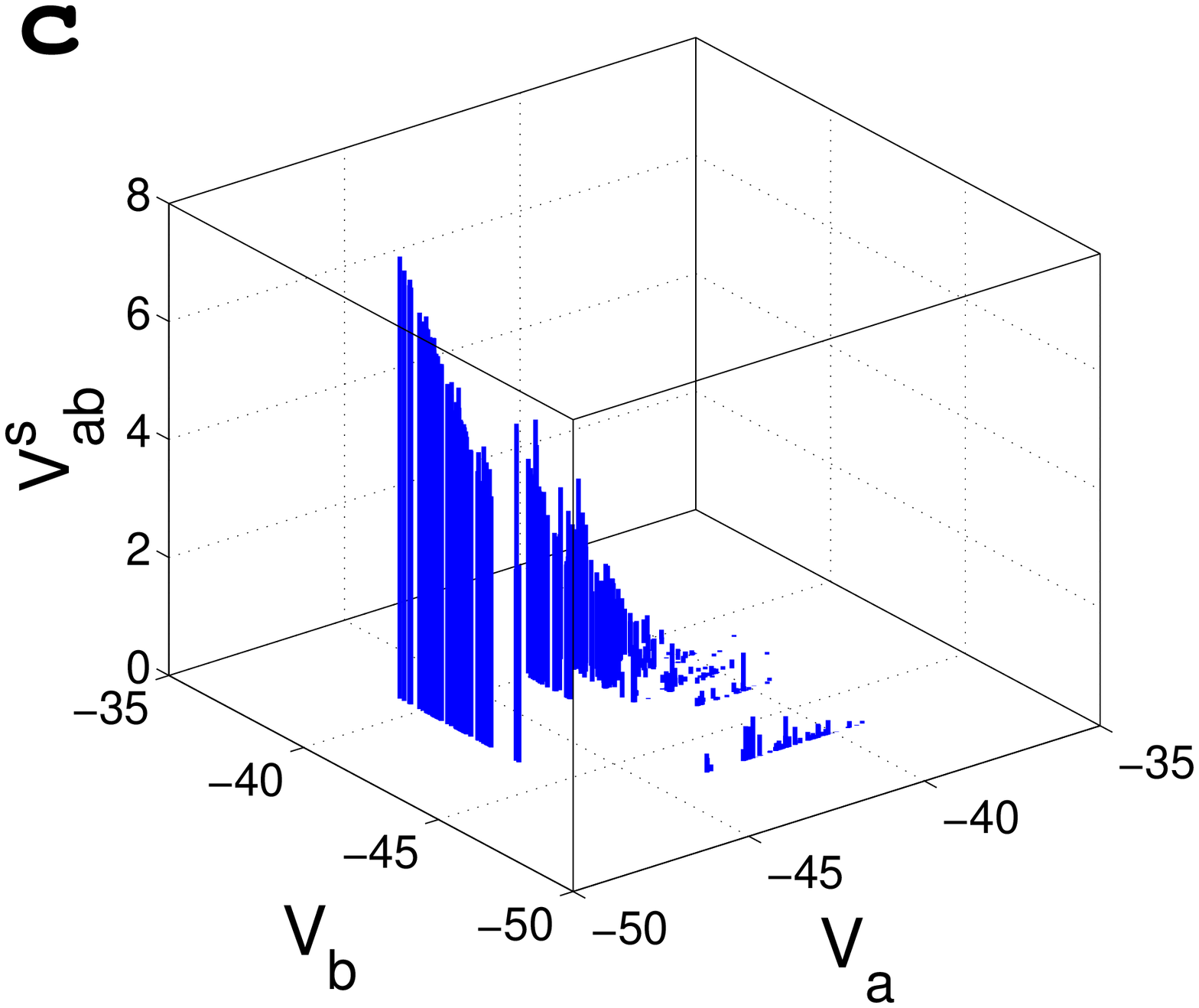}
\end{tabular}
\end{center}
\caption{ The essential PEL details, that go into the calculation of transition rates
[Eq.\eqref{eq.TransitionMatrix}], are shown. Panel {\bf a}: Histogram of IS energies. Panel {\bf b}: Dependence of the overall
curvature of a minimum on its energy. Panel {\bf c}: Heights of the lowest energy barriers
for going from minimum $a$ (with potential energy $V_a$) to minimum $b$ (with potential energy $V_b$).}
\label{fig.PESDetails}
\end{figure*}

\subsection{Diffusion constant and waiting times} \label{subsec.Diffusionconstant}

  The diffusion constant $D$ can be calculated from the mean square displacement $\langle R^2(t)\rangle$ using
$R_{ab}^2=N^{-1}\sum_{i=1}^N|\mathbf{r}_i^{(a)}-\mathbf{r}_i^{(b)}|^2$ in place of $\Phi_{ab}$ in
Eq.\eqref{eq.CorrelationFunction} ($\mathbf{r}_i^{(a)}$ denotes the position of the $i$-th particle in the $a$-th
minimum) and invoking the definition $D\equiv\lim_{t\rightarrow \infty}\frac{<R^2(t)>}{6t}=\lim_{t\rightarrow
\infty}\frac{1}{6}\frac{d <R^2(t)>}{dt}$. This definition would yield $D=0$ here as $D=\lim_{t\rightarrow \infty}\sum_{a,b,n} \lambda_n
(P^0_aP^0_b)^{1/2}R_{ab}^2 e_a^{(n)}e_b^{(n)} \exp{(-|\lambda_n|t)}=0$. In other words, $\langle R^2(t)\rangle\sim
t^{\alpha(t)}$ and $\lim_{t\rightarrow \infty}\alpha(t)=0$. Instead, we can define the diffusive regime as the
time window over which $\alpha(t) \simeq 1$ and then extract $D$ from the slope of the $\langle R^2(t)\rangle$ vs.~$t$ plot in this region.
This may not be useful for confined systems like the atomic cluster considered here, since such
a diffusive regime may be very short or
absent altogether, rendering $D$ to be an ill-defined quantity. Nevertheless, in this model we can define $D$
from the initial slope of the $\langle R^2(t)\rangle$ vs.~$t$ curve as the usual ballistic regime ($\langle
R^2(t)\rangle\propto t^2$),
seen for instance at very short times in MD simulations, is absent by definition. So, we may define the diffusion constant as
\begin{eqnarray}
D&\equiv&\lim_{t\rightarrow 0} \frac{d\langle R^2(t)\rangle}{dt}=\sum_{a,b} (P_a^0P_b^0)^{1/2}R_{ab}^2
\widetilde{W}_{ab}.\label{eq.DiffusionConstant}
\end{eqnarray}

The \emph{waiting time} in the basin of a minimum is the amount of time the system spends between an entry into and the subsequent exit from
the basin (i.e. during a single visit to the basin). The average of this quantity for the $a$-th minimum,
$\langle\tau_w(V_a)\rangle$,
can be calculated both from MD simulations and in the network model. In the master equation based model it is
straightforward to write $\langle\tau_w(V_a)\rangle$ as
\begin{eqnarray}
\langle\tau_w(V_a)\rangle&\equiv&\tau_a=-\frac{1}{W_{aa}}.
\end{eqnarray}
 On the other hand, if we consider an ergodic MD trajectory, then the amount of time the system spends in
$\mathcal{B}_a$ is $t_a\propto P_a^0$. Hence the total number of visits to $\mathcal{B}_a$ over the full
trajectory can be written as $v_a\equiv (t_a/\tau_a)\propto (P_a^0/\tau_a)$. Finally, the mean waiting time
averaged over the whole landscape would be
\begin{eqnarray}
\tau_w&=&\frac{\sum_a \tau_a v_a}{\sum_a v_a}=-\frac{1}{\sum_a P_a^0W_{aa}}. \label{eq.WaitingTime}
\end{eqnarray}
The above quantity can be related to $D$ [Eq.\eqref{eq.DiffusionConstant}] if $R_{ab}^2\simeq \bar{R}^2$ for
all pairs of minima, as $D\simeq \bar{R}^2\sum_{a\neq b}(P_a^0
P_b^0)^{1/2}\widetilde{W}_{ab}=-\bar{R}^2\sum_aP_a^0W_{aa}=\bar{R}^2/\tau_w$. Hence for this special case the
hopping between the basins becomes a random walk with a distribution of waiting times \cite{BDoliwa2}.

\section{Construction of the network for a 13-atom Morse cluster} \label{sec.NetworkMorseCluster}

 The Morse potential \cite{PMMorse} can be written in the form
\begin{eqnarray}
V=\sum_{i<j}V_{ij},~~~~~V_{ij}=e^{\rho(1-r_{ij}/r_e)}[e^{\rho(1-r_{ij}/r_e)}-2]\epsilon,
\label{eq.MorsePotential}
\end{eqnarray}
where $r_{ij}$ is the distance between atoms $i$ and $j$, $\epsilon$ and $r_e$ are the dimer well depth and
the equilibrium bond length, respectively - they simply scale the PEL without affecting its topology and can
conveniently be set to unity and used as the units of energy and distance. The parameter $\rho$ is a
dimensionless quantity that determines the range of the potential, with low values corresponding to long range. We
have taken $\rho=4$. This potential is widely used to model inter-atomic interactions in small atomic clusters or 
molecules \cite{DJWales1}. The reduced unit of time can be set to $(mr_e^2/\epsilon)^{1/2}$, $m$ being the atomic mass.

We follow more-or-less the same procedure (described briefly in Appendix \ref{app.BuildNetwork}) as that mentioned in
Ref.\cite{MAMiller2} for
building the network of minima and transition states. The network that we obtain consists of 138 minima and 230
transition states connecting them. We have not enforced any confining potential to prevent the cluster from
melting - in the temperature range of our interest, the particles are confined due to interactions among themselves
(we have discarded minima with maximum inter-particle separation more than 2.5). The minima and the transition
states around a particular minimum are identified by the values of their potential energy.
  
In Fig.\ref{fig.PESDetails} {\bf a}, the distribution of inherent structure energies is shown. We index the inherent structures in
ascending order of energy. The lowest lying minimum (the global minimum) is at $V_a=-46.635$ and after a
substantial gap there are three minima at $V_a\simeq -43.5$; after that, another perceptible gap is present and the rest of the
minima are closely spaced for $V_a \apgt -42.5$. Henceforth, we denote by $\mathcal{N}_f$ the full network and by $\mathcal{N}_r$
the network with the four lowest-lying minima removed. When we remove a particular minimum, all its edges and
minima that are connected to the rest of the network solely through this particular minimum get deducted from the
network as well. As a result, $\mathcal{N}_r$ contains 52 minima and 70 transition states.

The overall curvature at a minimum, $\mathcal{C}$ is obtained from the determinant of the Hessian matrix at the
minimum i.e.~$\mathcal{C}=\mathrm{Det}(\mathbf{H})$. Here the deeper basins are narrower, as is evident from
Fig.\ref{fig.PESDetails} {\bf b}. Also the number of barriers and barrier heights increase with decreasing
IS energies (Fig.\ref{fig.PESDetails} {\bf c}).

In the next section we describe the results of the master equation based calculations carried out with the
networks $\mathcal{N}_f$ and $\mathcal{N}_r$. While the relaxation dynamics in $\mathcal{N}_f$ is quite similar to
that of strong glass formers, the dynamics in $\mathcal{N}_r$ shows strong resemblance to that of fragile ones.
Although $\mathcal{N}_r$ is a part of $\mathcal{N}_f$, the relaxation of the system restricted in
this part of the configuration space becomes qualitatively very different from the global dynamics in $\mathcal{N}_f$.

The procedure of searching for transition states (Appendix \ref{app.BuildNetwork}) starting at a minimum and moving along
directions of successively larger eigenvalues of the Hessian matrix is not very efficient \cite{TFMiddleton} and most of
the time, one ends up at the same transition state. Hence we get only 3 barriers around a
minimum on the average. Although the details of the connectivity of the network may matter for finer details, our main 
objective in this work has been to realize the typical characteristics of glassy
dynamics with a minimal set of minima and transition states. Hence we have not paid much attention to building the
network very accurately so as to represent the actual system. Nevertheless, the network model seems to capture
the main features of the complex long-time dynamics quite accurately, as verified through MD simulations (Section
\ref{sec.MD}).

\begin{figure}
\begin{center}
\begin{tabular}{cc}
\includegraphics[width=4cm]{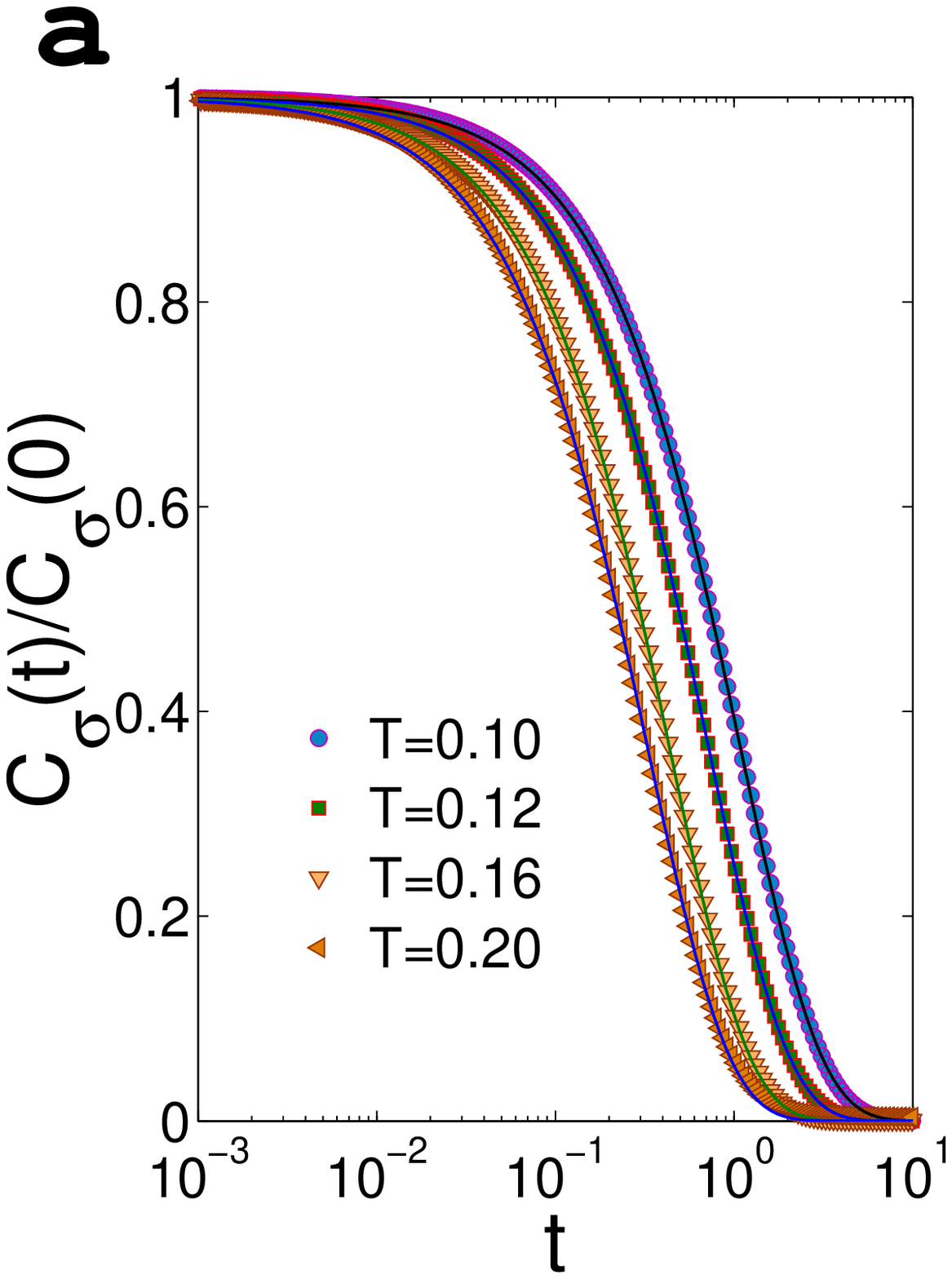} &
\includegraphics[width=4cm]{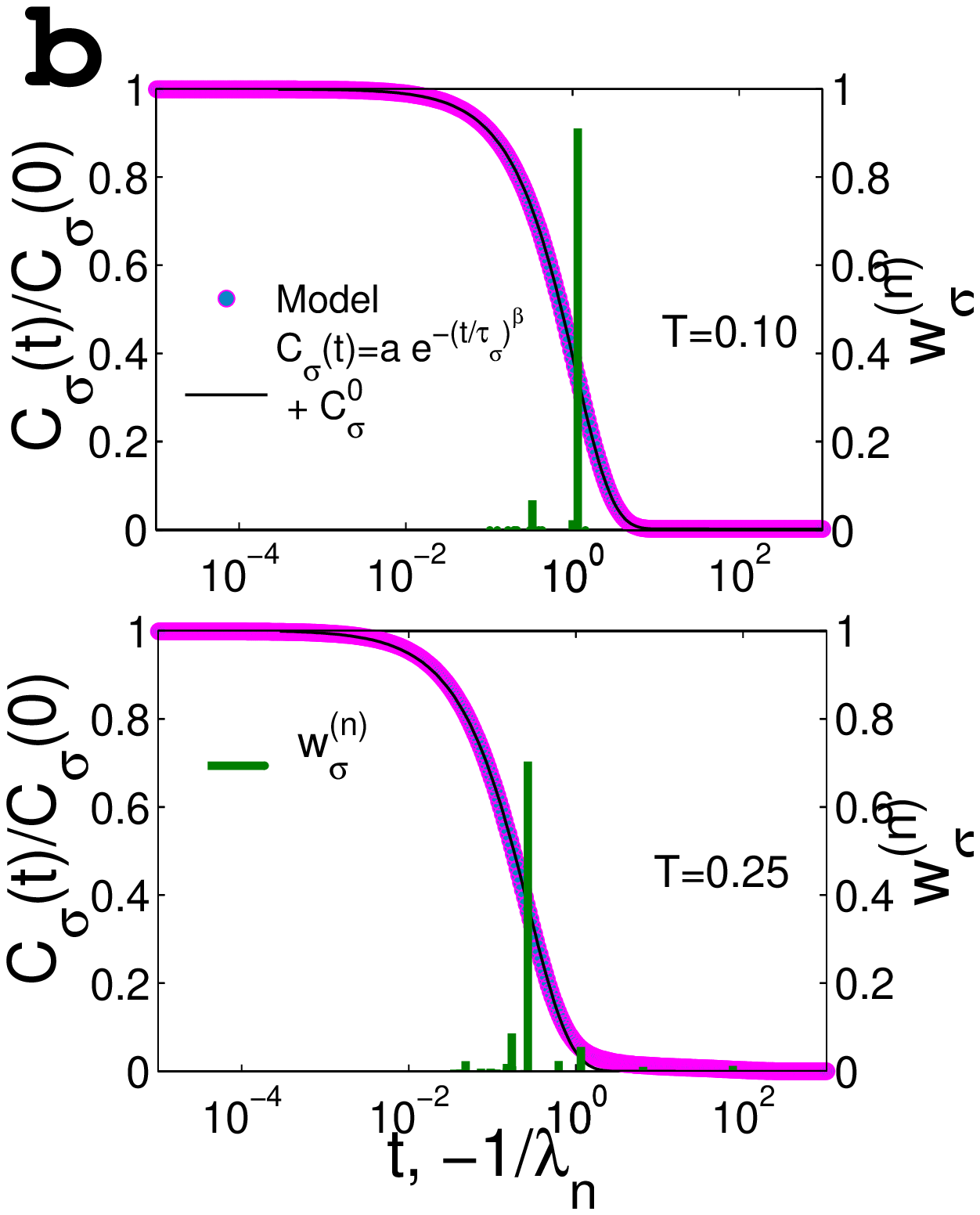}\\
\includegraphics[width=4cm]{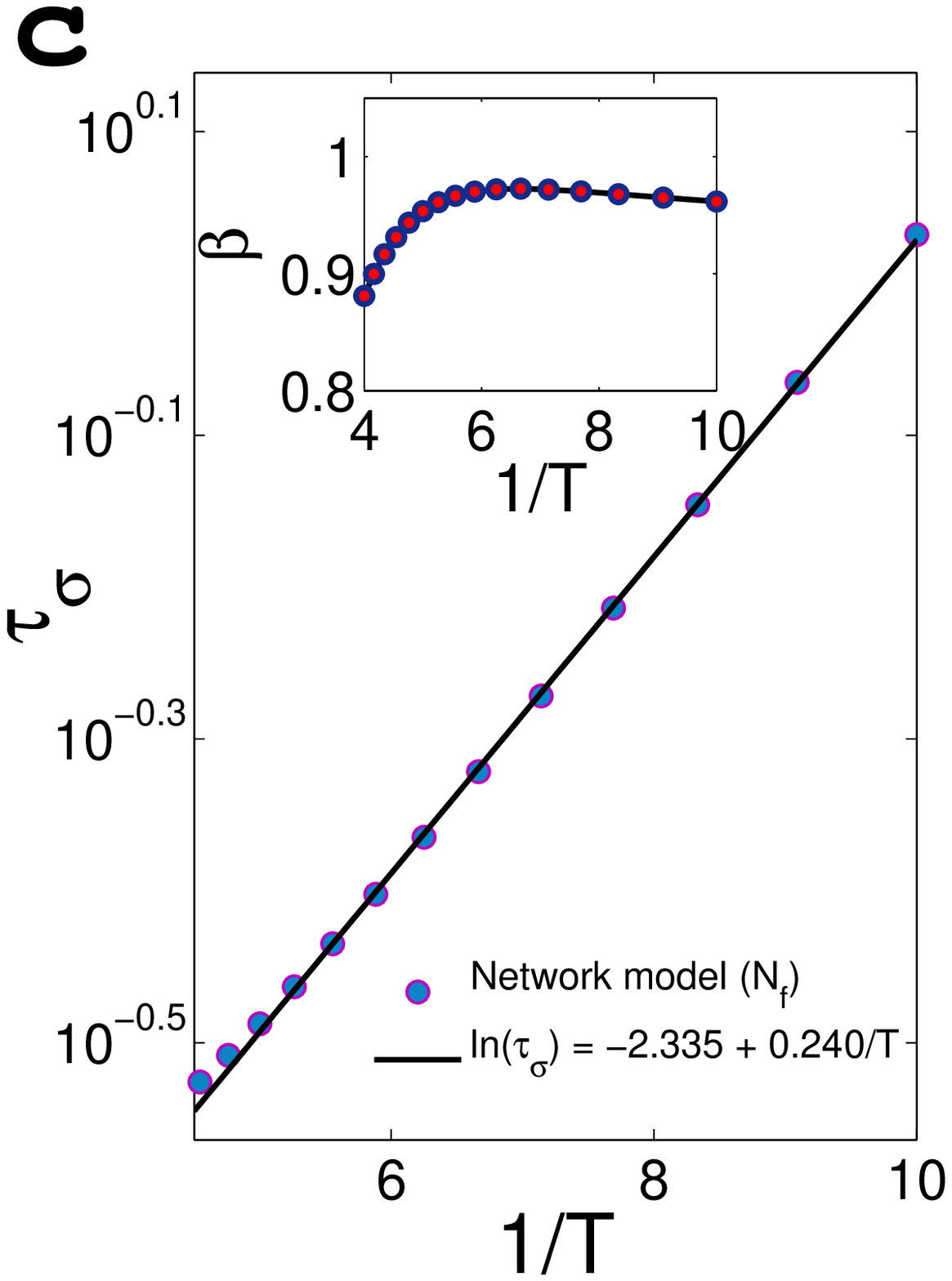}&
\includegraphics[width=4cm]{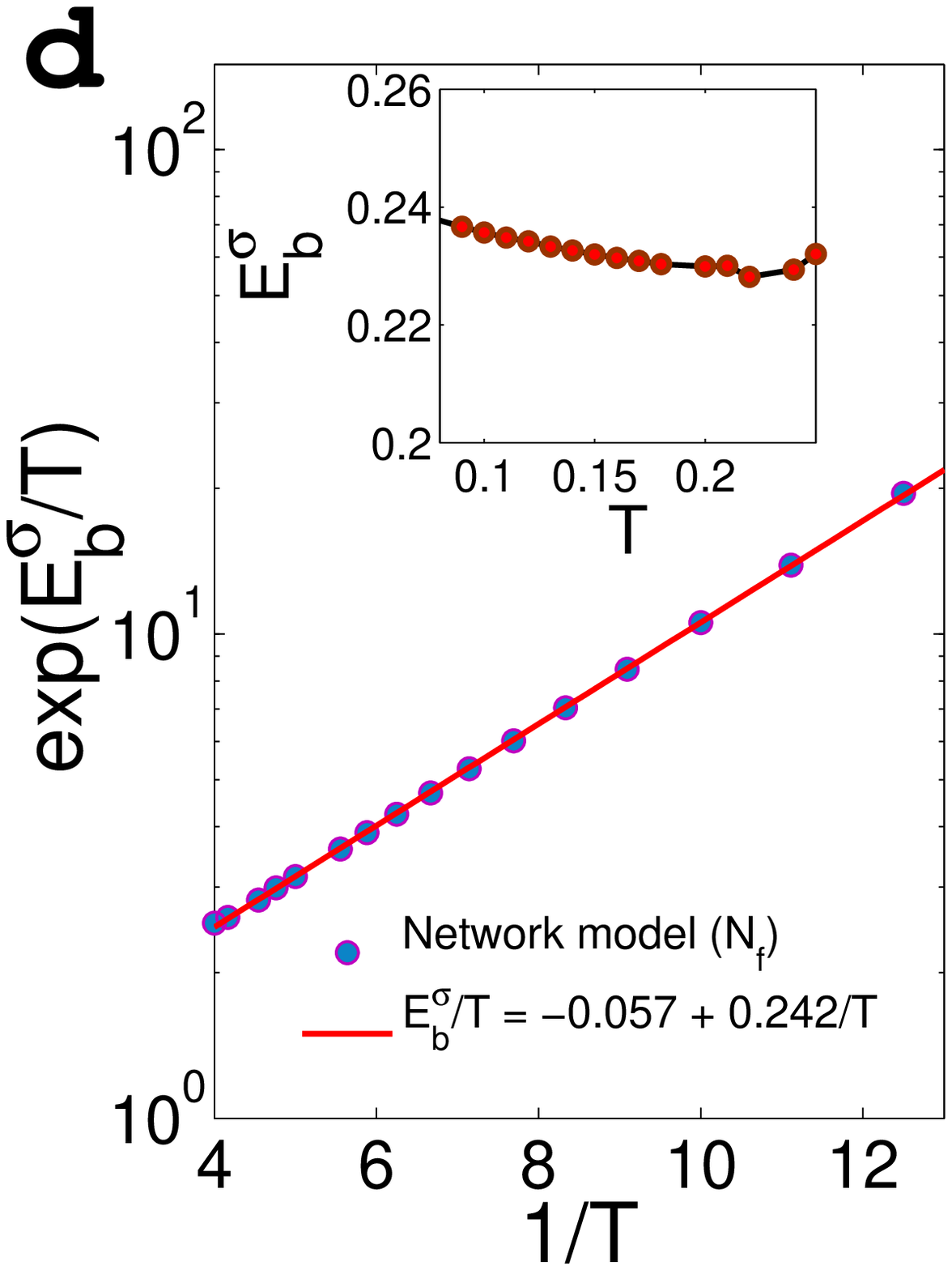} 
\end{tabular}
\end{center}
\caption{Stress autocorrelation function for the network $\mathcal{N}_f$ (Section
\ref{sec.NetworkMorseCluster}). Panel {\bf a}: $C_\sigma(t)$ [normalized by dividing with $C_\sigma(0)$] is
shown for various $T$. Panel {\bf b}: Decay of $C_\sigma(t)$ is determined by one dominant relaxation mode $n$ over
the entire temperature range of interest, as is evident from the  plot of $w_\sigma^{(n)}$ vs.~$|\lambda_n|^{-1}$
[Eq.\eqref{eq.ModeWeight}] for $T=0.10,~0.25$. Panel {\bf c}: Arrhenius plot for $\tau_\sigma(T)$ [extracted by fitting
$C_\sigma(t)$ with the KWW form of Eq.\eqref{eq.KWW}] i.e.~$\ln \tau_\sigma$ vs.~$1/T$. The effective barrier
$E_b^\sigma$ is obtained by fitting the data to the Arrhenius form [Eq.\eqref{eq.Arrhenius}]. The stretching
exponent $\beta\simeq 1$ (inset) confirms the simple exponential nature of the decay. Panel {\bf d}: The
estimate of Eq.\eqref{eq.EffectiveBarrier} (Appendix \ref{app.ActivationEnergy}) for $E_b^\sigma$ agrees very
well with the value of $E_b^\sigma=0.24$ obtained from the Arrhenius fit shown in panel {\bf c}.}
\label{fig.StressAutoCorrelation_Nf}
\end{figure}

\section{Strong and fragile behavior in the network model}
\label{sec.DynamicsMorseCluster}

Here we describe the results for the stress-stress autocorrelation function $C_\sigma(t)$, as described in
Sec.\ref{subsec.Structuralrelaxation}, for the Morse cluster.

Fig.\ref{fig.StressAutoCorrelation_Nf} {\bf a} shows $C_\sigma(t)$ for the network $\mathcal{N}_f$ at four different
temperatures. The decay of the correlation function is very well described by a single exponential [the stretching
exponent of Eq.\eqref{eq.KWW}, $\beta\simeq 1$] over the entire temperature range, as is evident from the inset of
Fig.\ref{fig.StressAutoCorrelation_Nf} {\bf c}. The origin of this simple Debye-like relaxation can be
attributed to the presence of a single relaxation mode with a large weight $w_\sigma^{(n)}$
[Eq.\eqref{eq.ModeWeight}] as shown in Fig.\ref{fig.StressAutoCorrelation_Nf} {\bf b}. The corresponding relaxation 
time $\tau_\sigma$ extracted by fitting $C_\sigma(t)$ with Eq.\eqref{eq.KWW} follows an Arrhenius temperature
dependence (Fig.\ref{fig.StressAutoCorrelation_Nf} {\bf c}), i.e.
\begin{eqnarray}
\tau_\sigma=\tau_\sigma^0\exp{\left(\frac{E_b}{T}\right)} \label{eq.Arrhenius}
\end{eqnarray} 
In $\mathcal{N}_f$, the dynamics is mainly
governed by the barriers connecting the global minima at $V_a\simeq -46.6$ with the next three lowest lying
minima at $V_a\simeq -43.5$ (see Section \ref{sec.NetworkMorseCluster}). 

The Arrhenius fit to $\tau_\sigma$
vs. $T$ curve yields an effective activation barrier $E_b^\sigma=0.24$ which one can easily identify with the barrier 
for going from one of the minima at $V_a\simeq -43.5$ to the global minimum. The estimate for
$E_b^\sigma$ (Fig.\ref{fig.StressAutoCorrelation_Nf} {\bf d}) obtained from Eq.\eqref{eq.EffectiveBarrier} 
(Appendix \ref{app.ActivationEnergy}) agrees very well with
the above value. In contrast, we find that the barrier for going from the global minimum to the next lowest lying
minimum determines the effective activation barrier that appears in the $T$-dependence of the
mean waiting time $\tau_w$ [Eq.\eqref{eq.WaitingTime}]. This is a natural consequence of the fact that the
global minimum, being much lower in potential energy with respect to the other minima in this case, possesses
almost all the Boltzmann weight. As a result, the relaxation to equilibrium (decay of correlation) is entirely
dictated by the relaxation paths from other parts of the PEL to the global minimum. The mean
waiting time $\tau_w$, on the other hand, is decided by the escapes from the global minimum over the barriers
surrounding it as the system spends most of the time in the basin of the global minimum.  

The above observations suggest a trivial route for realizing strong behavior, namely
dynamics governed by a fixed set of barriers surrounding a very deep inherent structure (or a set of
inherent structures with very similar potential energies in a more general case) possessing most of the 
Boltzmann occupation probability. Keeping this fact in mind we construct the network $\mathcal{N}_r$ by
removing a few deep minima so that a larger number of minima figure in the relaxation to equilibrium due to 
comparable Boltzmann
weights in the activated regime and many different barriers contribute to the relaxation process.

\begin{figure}
\begin{center}
\begin{tabular}{cc}
\includegraphics[width=4cm]{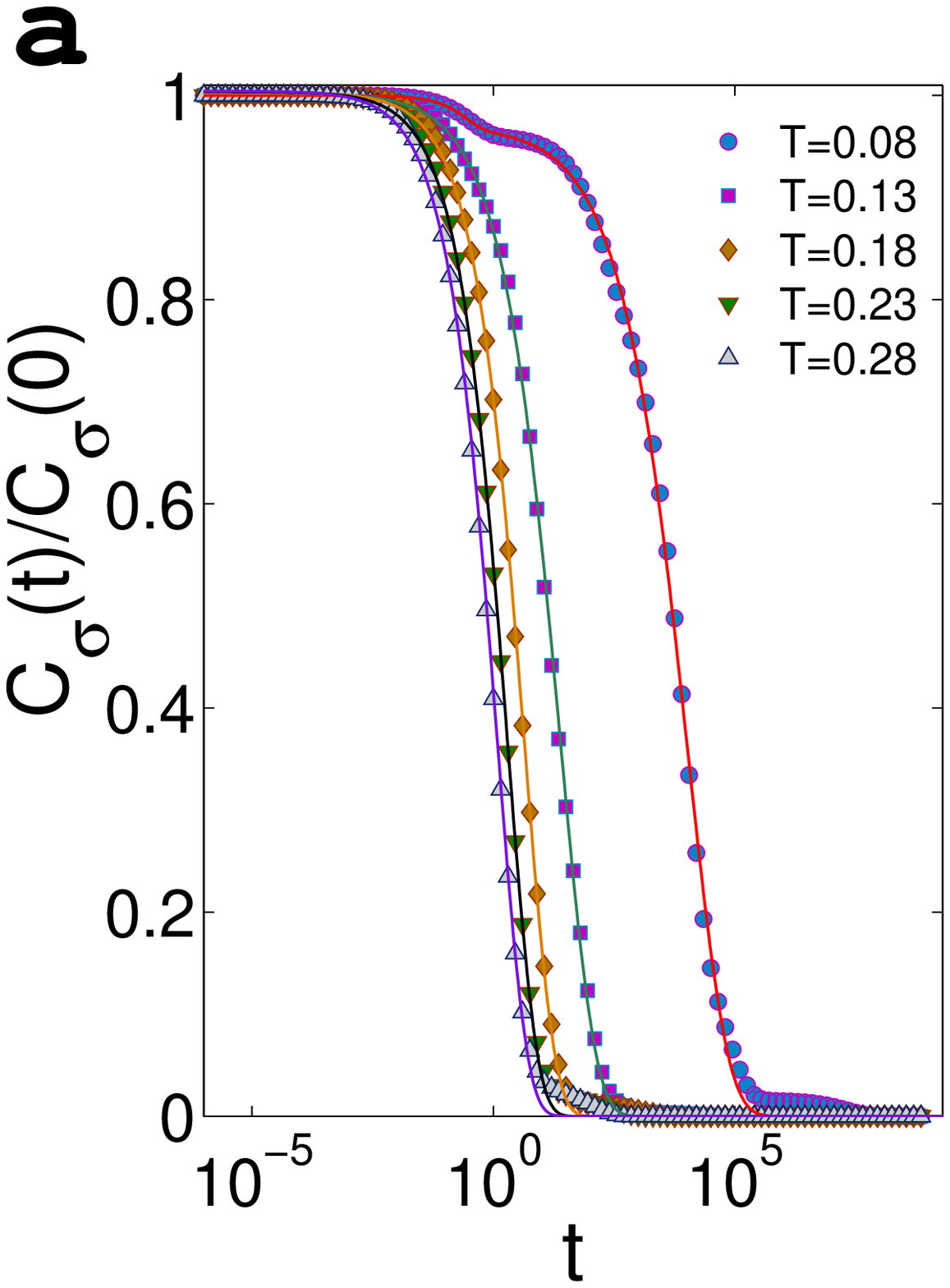} &
\includegraphics[width=4cm]{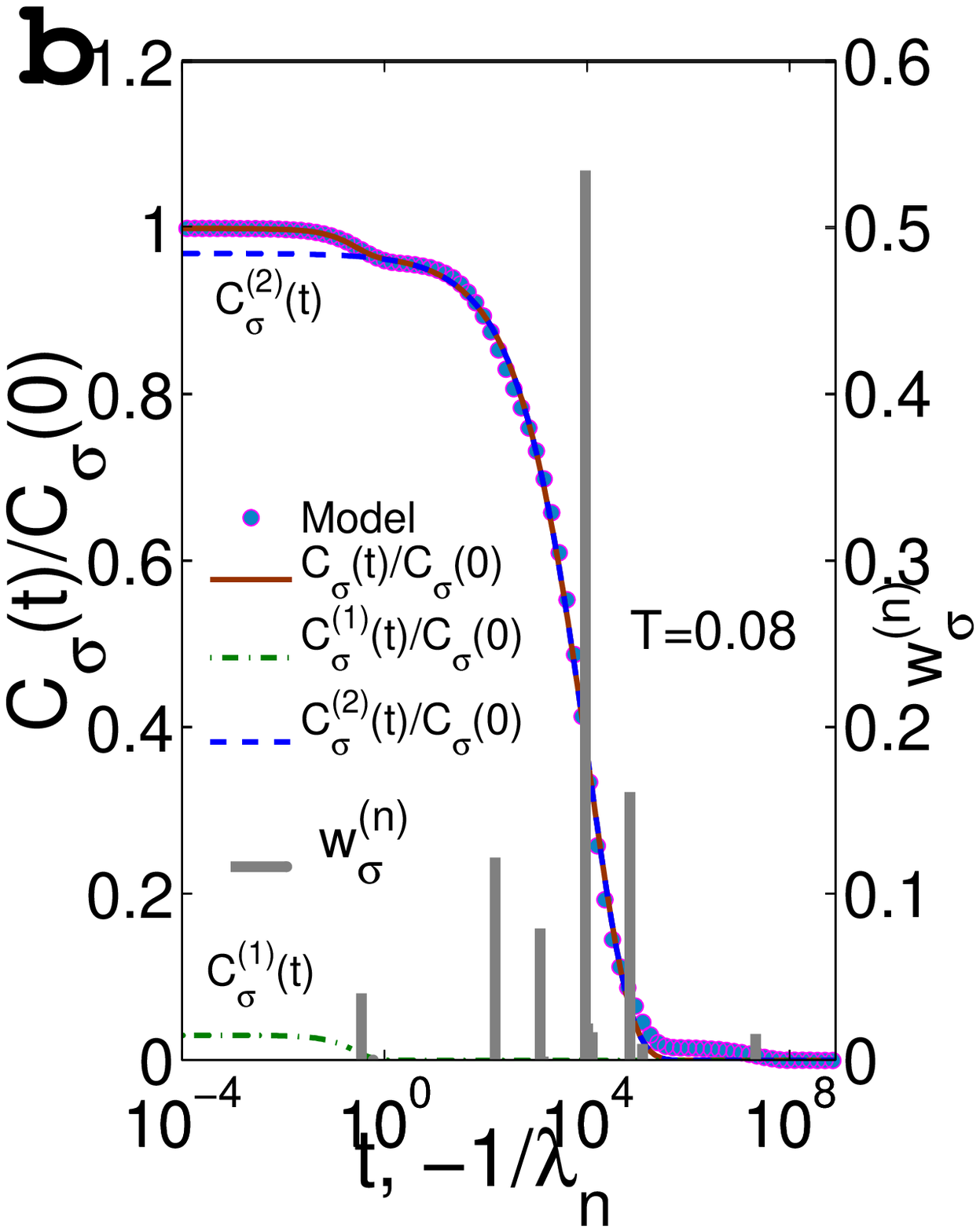}\\
\includegraphics[width=4cm]{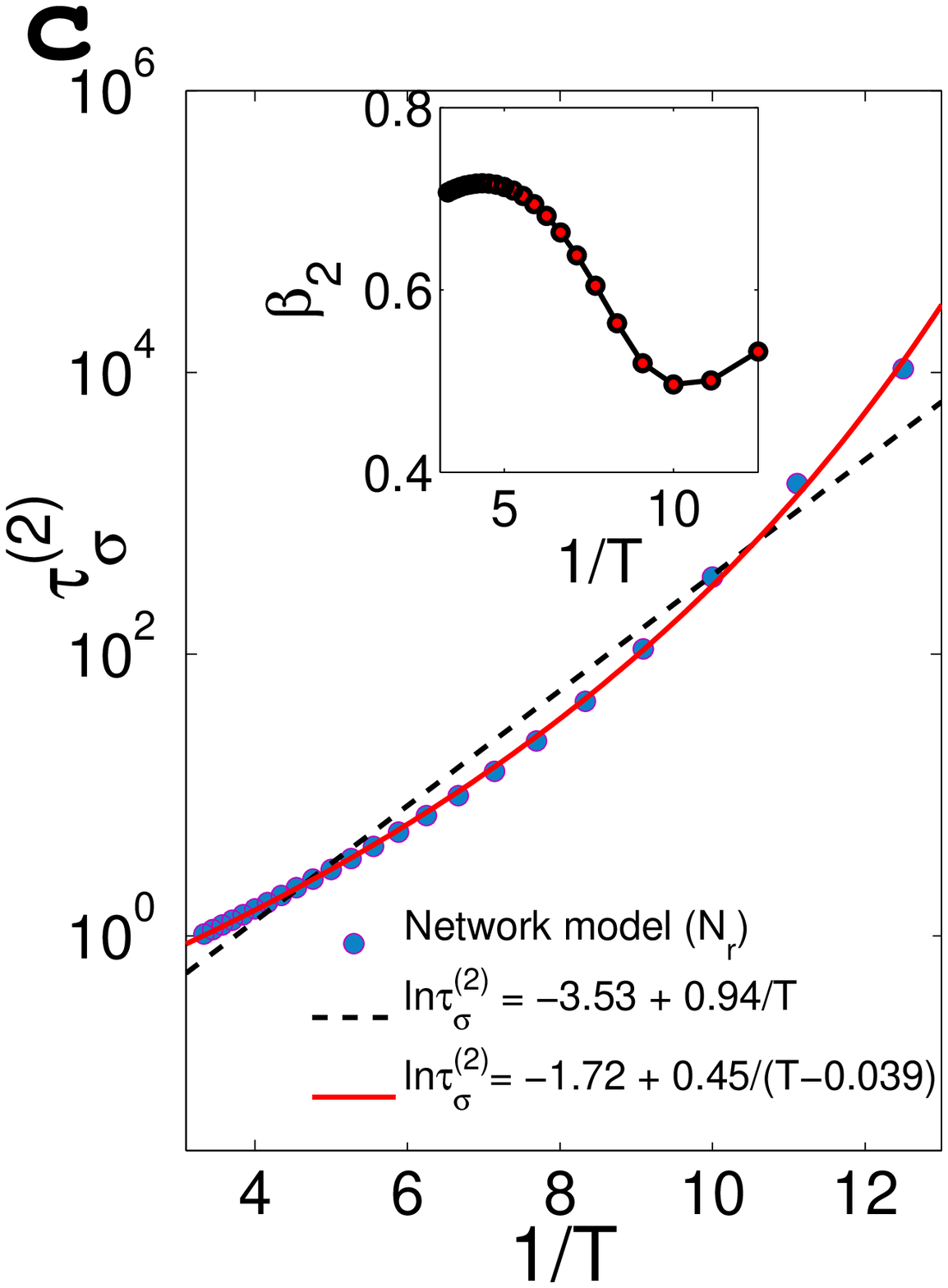}&
\includegraphics[width=4cm]{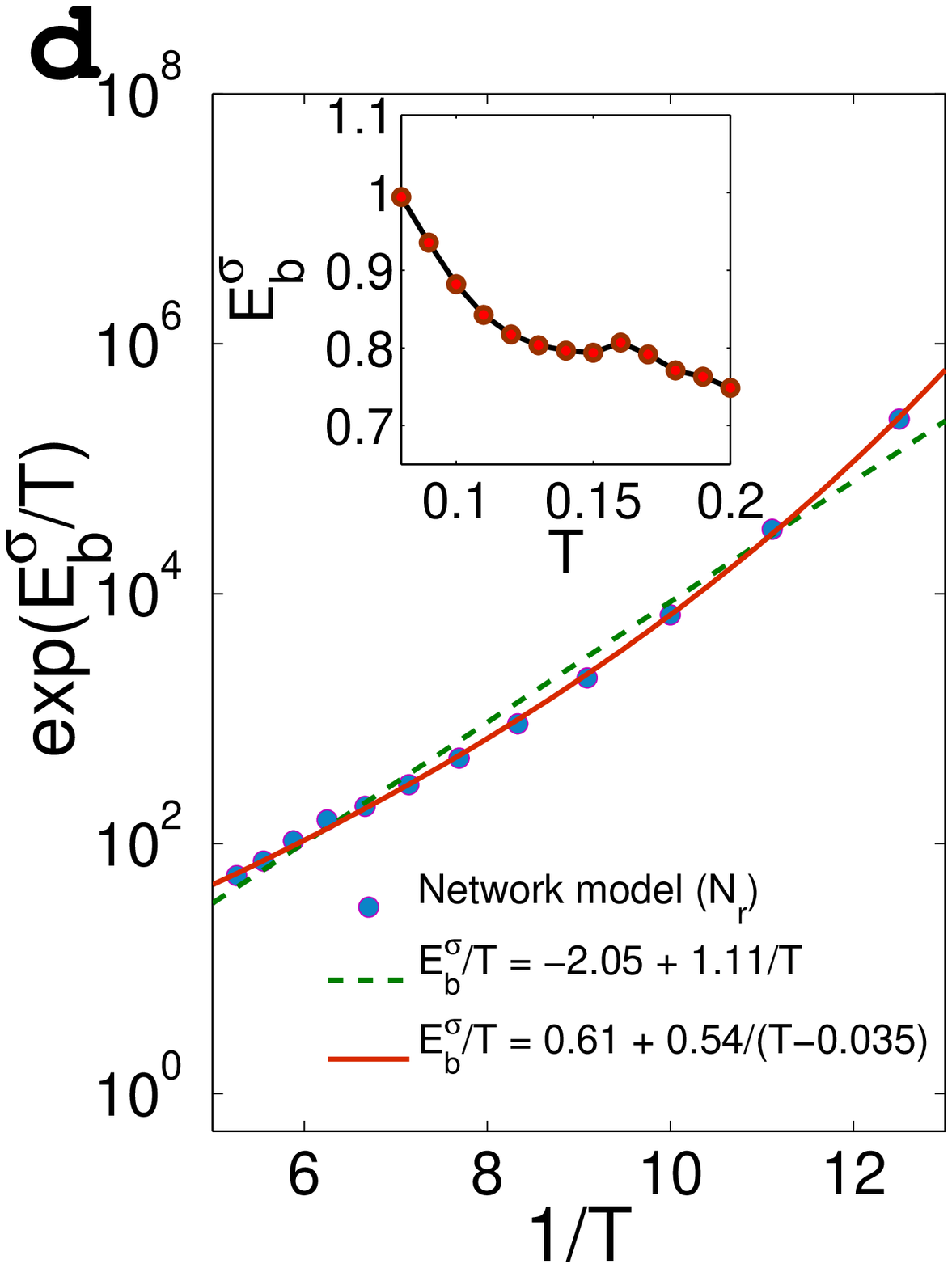}
\end{tabular}
\end{center}
\caption{Stress autocorrelation function in $\mathcal{N}_r$. Panel {\bf a}: $C_\sigma(t)$ is shown for five $T$
values. Panel {\bf b}: Fit to a sum of two stretched exponentials (Eq.\eqref{eq.KWW}, see the text for
details) at $T=0.08$. The quantity $w_\sigma^{(n)}$ exhibits the presence of (well-separated) multiple timescales even at
the very low temperature $T=0.08$. Panel {\bf c}: The deviation from the simple Arrhenius form is evident from the plot
of $\ln\tau_\sigma^{(2)}$ vs.~$1/T$. Fits to both the Arrhenius form
[Eq.\eqref{eq.Arrhenius}] and the VFT
form [Eq.\eqref{eq.VFT}] are shown. It is clear that the VFT form provides a good fit. {\bf Inset}: The exponent $\beta_2$ is much less than 1, 
specially at low temperatures, showing the non-exponential nature of the relaxation. Panel {\bf d}, The effective
barrier $E_b^\sigma(T)$ obtained from Eq.\eqref{eq.EffectiveBarrier} agrees reasonably well with the
estimate deduced from the Arrhenius and VFT fits. For instance, the VFT fit to $\exp{(E_b^\sigma/T)}$
vs.~$1/T$ yields values for the parameters $B_\sigma$ and $T_0^\sigma$ [Eq.\eqref{eq.VFT}] that are similar to those
obtained in panel {\bf c}.}
\label{fig.StressAutoCorrelation_Nr}
\end{figure}

Fig.\ref{fig.StressAutoCorrelation_Nr} {\bf a} exhibits $C_\sigma(t)$ calculated for $\mathcal{N}_r$. We
observe a two-stage, non-exponential relaxation in this case. This, again, can be understood from the values of
$w_\sigma^{(n)}$ (Fig.\ref{fig.StressAutoCorrelation_Nr} {\bf b}). The profile of $C_\sigma(t)$ is well fitted with
a sum of two stretched exponentials,
i.e.~$C_\sigma(t)=C_\sigma^{(1)}(t)+C_\sigma^{(2)}(t)=C_\sigma^{(1)}(0)e^{-(t/\tau_\sigma^{(1)})^{\beta_1}}+
C_\sigma^{(2)}(0)e^{-(t/\tau_\sigma^{(2)})^{\beta_2}}$.
We plot the longer of the two relaxation times, $\tau_\sigma^{(2)}$, in Fig.\ref{fig.StressAutoCorrelation_Nr} {\bf
c} along with the exponent $\beta_2$ in the inset. The deviation from simple exponential behavior is evident.
The temperature dependence of $\tau_\sigma^{(2)}$ exhibits marked deviation from the simple Arrhenius behavior.
Rather, the Vogel-Fulcher-Tammann (VFT) form \cite{HVogel,GSFulcher,GTammann},
\begin{eqnarray}
\tau_\sigma=\tau_\sigma^0\exp{\left(\frac{B_\sigma}{T-T_0^\sigma}\right)}, \label{eq.VFT} 
\end{eqnarray}
frequently used for fragile glass formers,
yields a much better representation of the data for $\tau_\sigma^{(2)}$. The \emph{fragile} nature of
the $\ln \tau_\sigma^{(2)}$ vs.~$1/T$ curve is well reproduced in Fig.\ref{fig.StressAutoCorrelation_Nr} {\bf
d} by the effective temperature dependent barrier $E_b^\sigma(T)$ calculated from Eq.\eqref{eq.EffectiveBarrier}.   

\begin{figure*}
\begin{center}
\begin{tabular}{c}
\includegraphics[width=16cm]{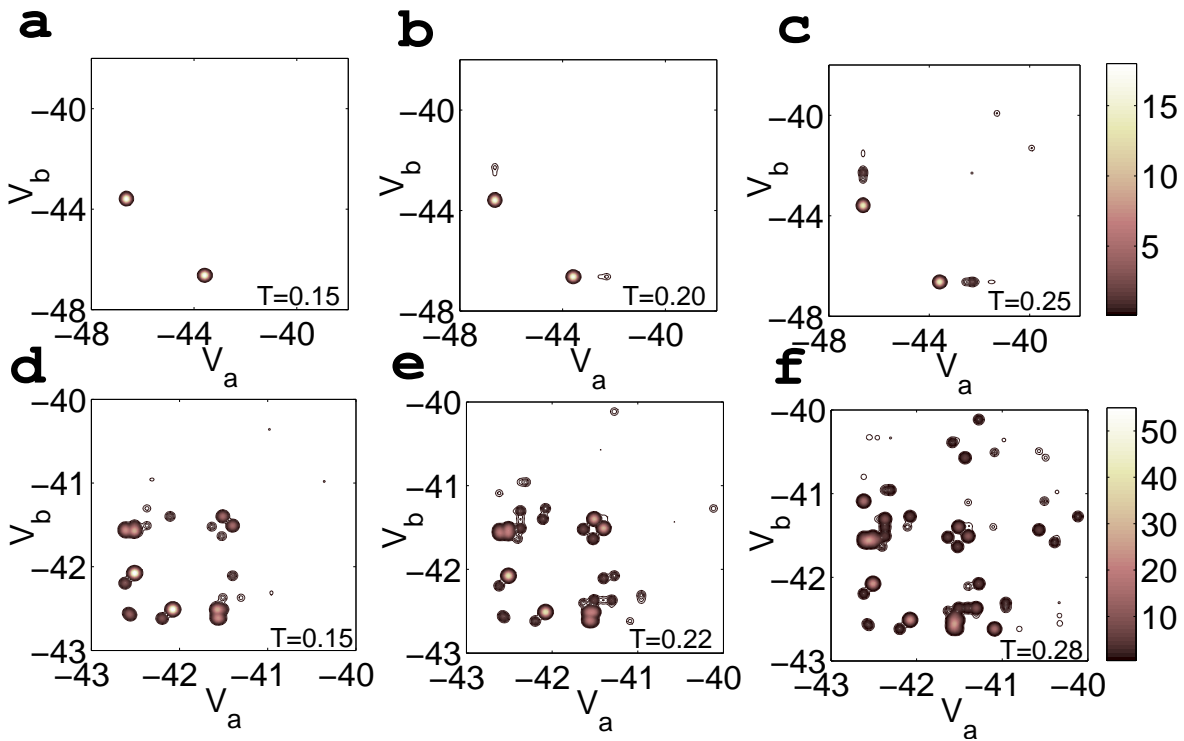}
\end{tabular}
\end{center}
\caption{ Color maps (see the color bars) for $\mathcal{W}_{ab}$ [Eq.\eqref{eq.MinimaPairWeight}] in the $(V_a,V_b)$ plane for
$\mathcal{N}_f$ at three temperatures, $T=0.15$ (panel {\bf a}), $T=0.20$ (panel {\bf b}), and $T=0.25$ (panel {\bf c}), and for
$\mathcal{N}_r$ at $T=0.15$ (panel {\bf d}), $T=0.22$ (panel {\bf e}), and $T=0.28$ (panel {\bf f}).}
\label{fig.MinimaPairWeight}
\end{figure*}

A useful visualization and understanding of the observed \emph{strong} behavior for the PEL $\mathcal{N}_f$
and \emph{fragile} behavior for $\mathcal{N}_r$ can be achieved by looking at the quantity $\mathcal{W}_{ab}$
defined in Eq.\eqref{eq.MinimaPairWeight} and the subsequent paragraph of Appendix
\ref{app.ActivationEnergy}. This quantity, $\mathcal{W}_{ab}$, can be thought of as the weight with which 
the edges between the nodes $a$ and $b$, or in other words, the elementary jumps over the barriers connecting
minima $a$ and $b$ appears in the relaxation process through various relaxation channels or modes $n$. In
Figs.\ref{fig.MinimaPairWeight} {\bf a}, {\bf b} and {\bf c} we have shown the color maps of $\mathcal{W}_{ab}$ 
for $\mathcal{N}_f$ at three temperatures. In these plots, the coordinates $(V_a,V_b)$ represent the 
barrier between minima $a$ and $b$ and the color
corresponds to the value of $\mathcal{W}_{ab}$ at $(V_a,V_b)$ (we have used a small broadening for the purpose of 
visualization). It is clear from the plots that only a few barriers surrounding the global
minimum contribute substantially to the relaxation process and over the entire temperature range, the peaks
remain at nearly the same positions. On the contrary for $\mathcal{N}_r$, in 
Figs. \ref{fig.MinimaPairWeight} {\bf d}, {\bf e} and {\bf f}, many barriers figure in the activated
relaxation and the picture changes substantially with increasing temperature, as more and more barriers start
to play a role in the relaxation process. This feature can be considered as the trademark of
\emph{fragile} dynamics.
  
We have observed similar characteristics of \emph{strong} and \emph{fragile} dynamics for the relaxation time
$\tau_\delta$ associated with the overlap function $C_\delta(t)$, defined in Section
\ref{subsec.Structuralrelaxation}, obtained from both the network model calculation and MD simulations.
We report these results in the next section. 

\begin{figure}
\begin{center}
\begin{tabular}{cc}
\includegraphics[width=4cm]{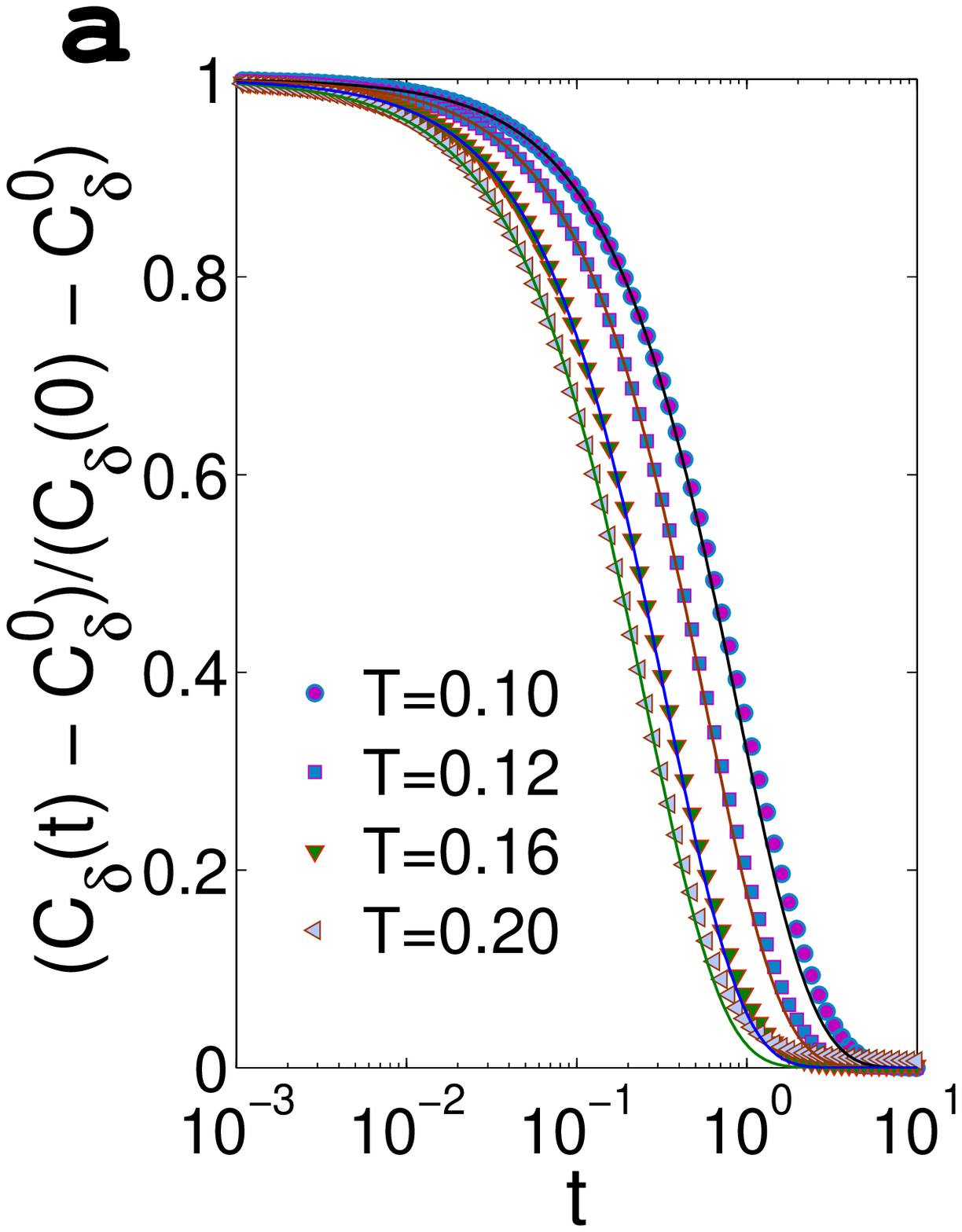} &
\includegraphics[width=4cm]{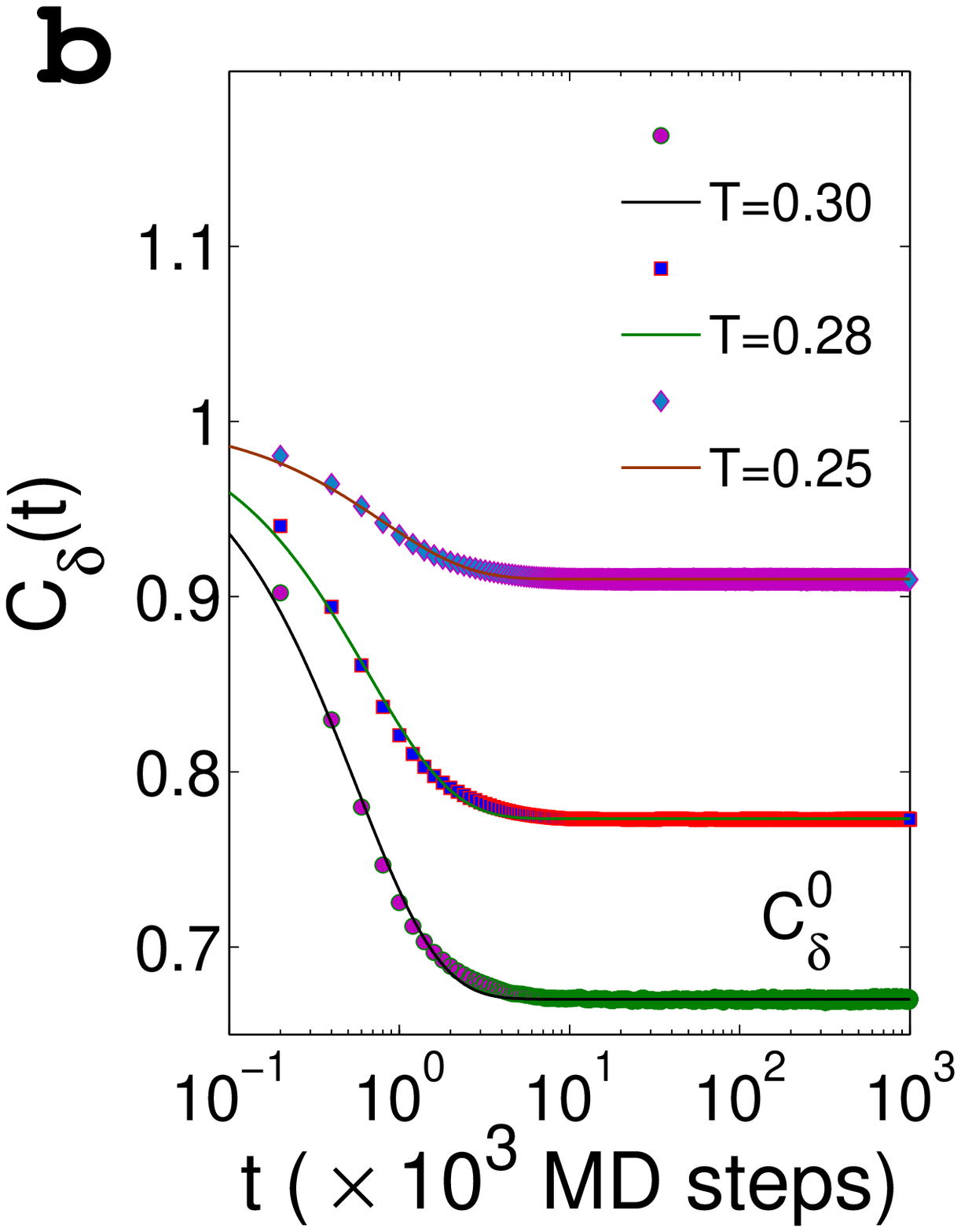}\\
\includegraphics[width=4cm]{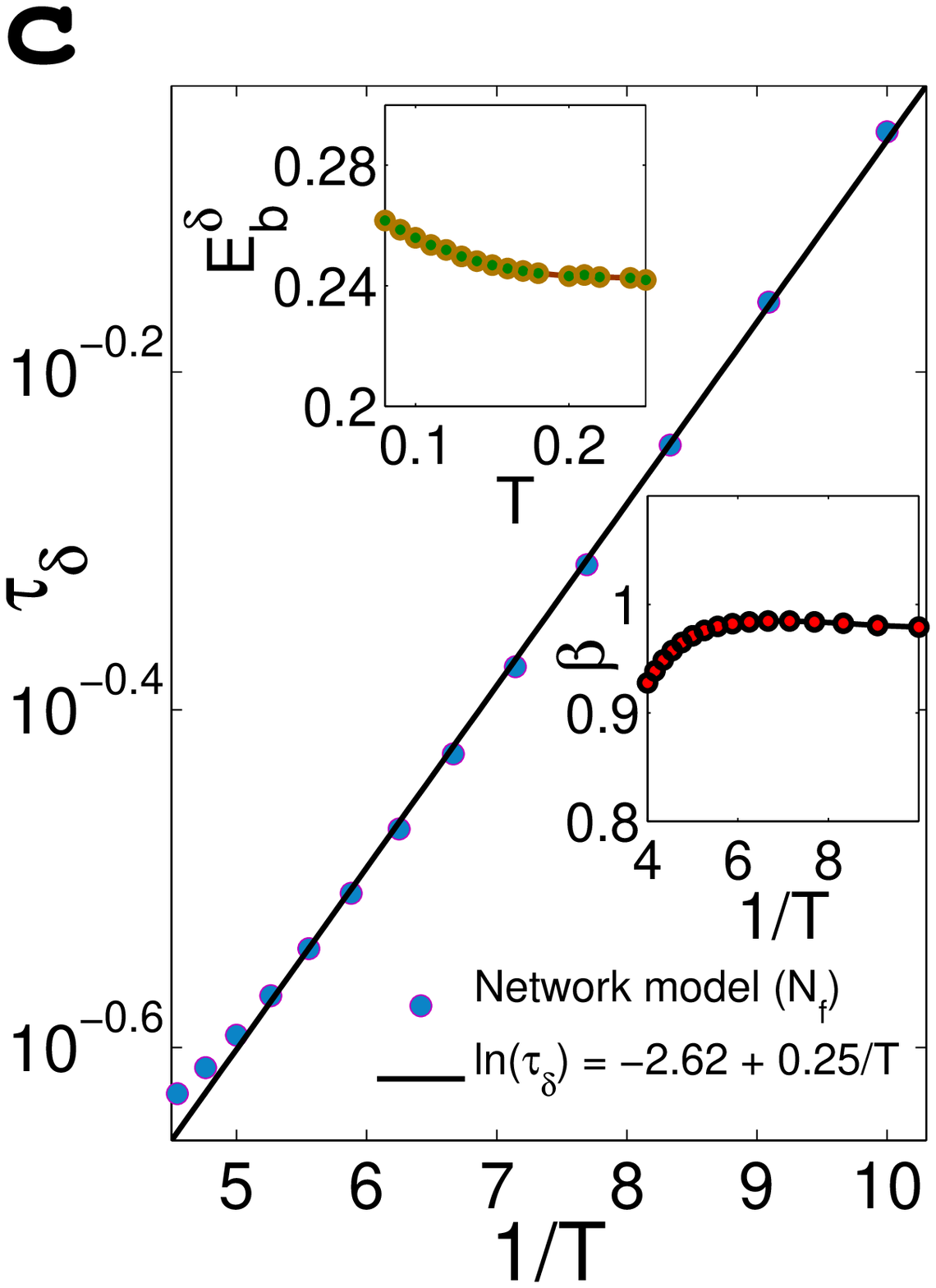}&
\includegraphics[width=4cm]{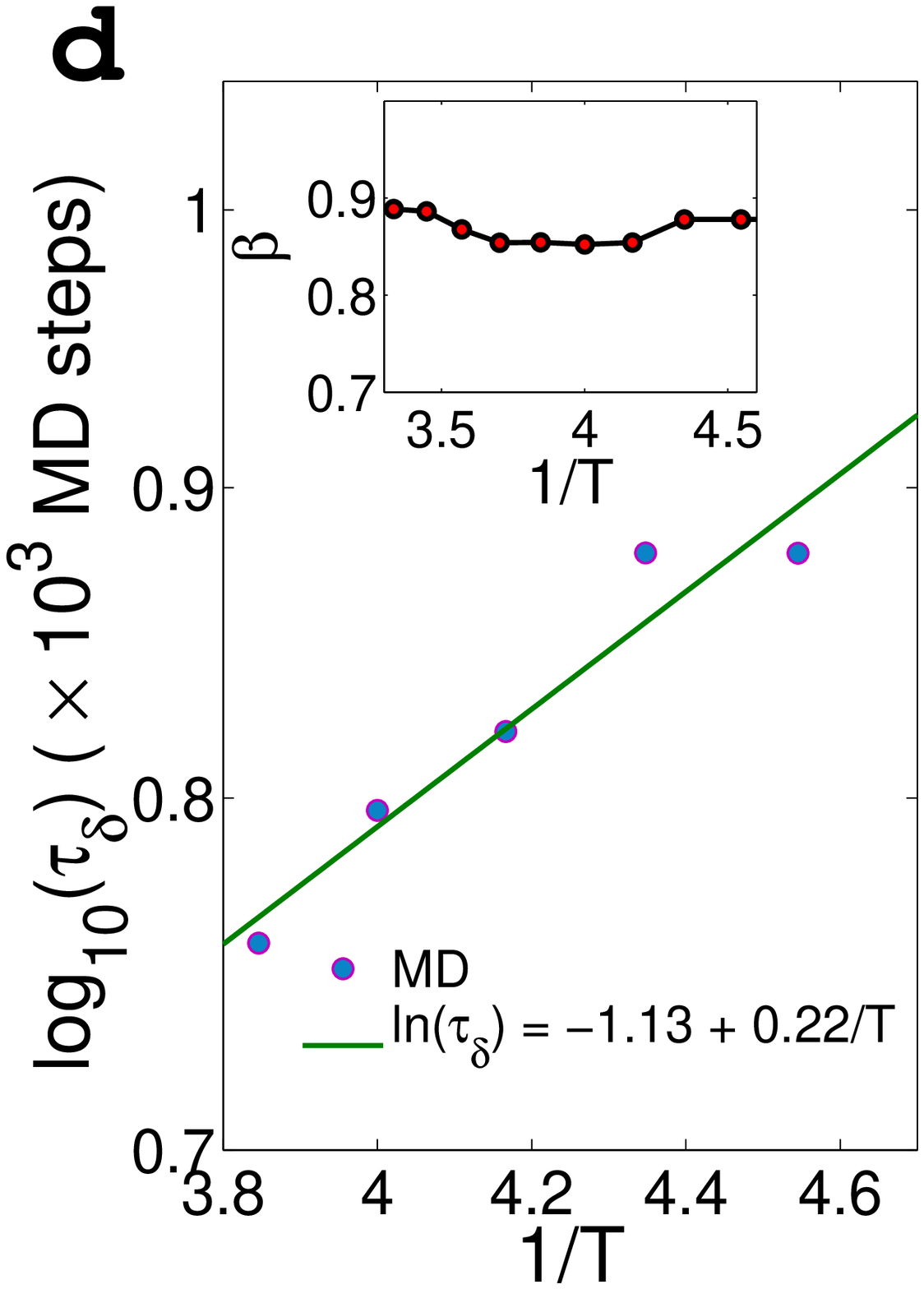}
\end{tabular}
\end{center}
\caption{Overlap function $C_\delta(t)$ for the network $\mathcal{N}_f$. Panel {\bf a}:
$(C_\delta(t)-C_\delta^0)/(C_\delta(0)-C_\delta^0)$ obtained from
the network model for four temperatures. Panel {\bf b}: MD results for $C_\delta(t)$. Panel {\bf c}: Arrhenius plot for
$\tau_\delta(T)$ deduced in {\bf a} from KWW fits [Eq.\eqref{eq.KWW}]. The top and bottom insets show, respectively, the estimates of
$E_b^\delta(T)$ from Eq.\eqref{eq.EffectiveBarrier} and the stretching exponent $\beta$ for the KWW fit to
$C_\delta(t)$ in {\bf a}. Panel {\bf d}: Estimates of $\tau_\delta$ can be obtained accurately only for a narrow
temperature range in MD, as discussed in the text. The exponent $\beta$, extracted from {\bf b}, is shown in
the inset.}
\label{fig.OverlapCorrelation_Nf}
\end{figure}

\begin{figure}
\begin{center}
\begin{tabular}{cc}
\includegraphics[width=4cm]{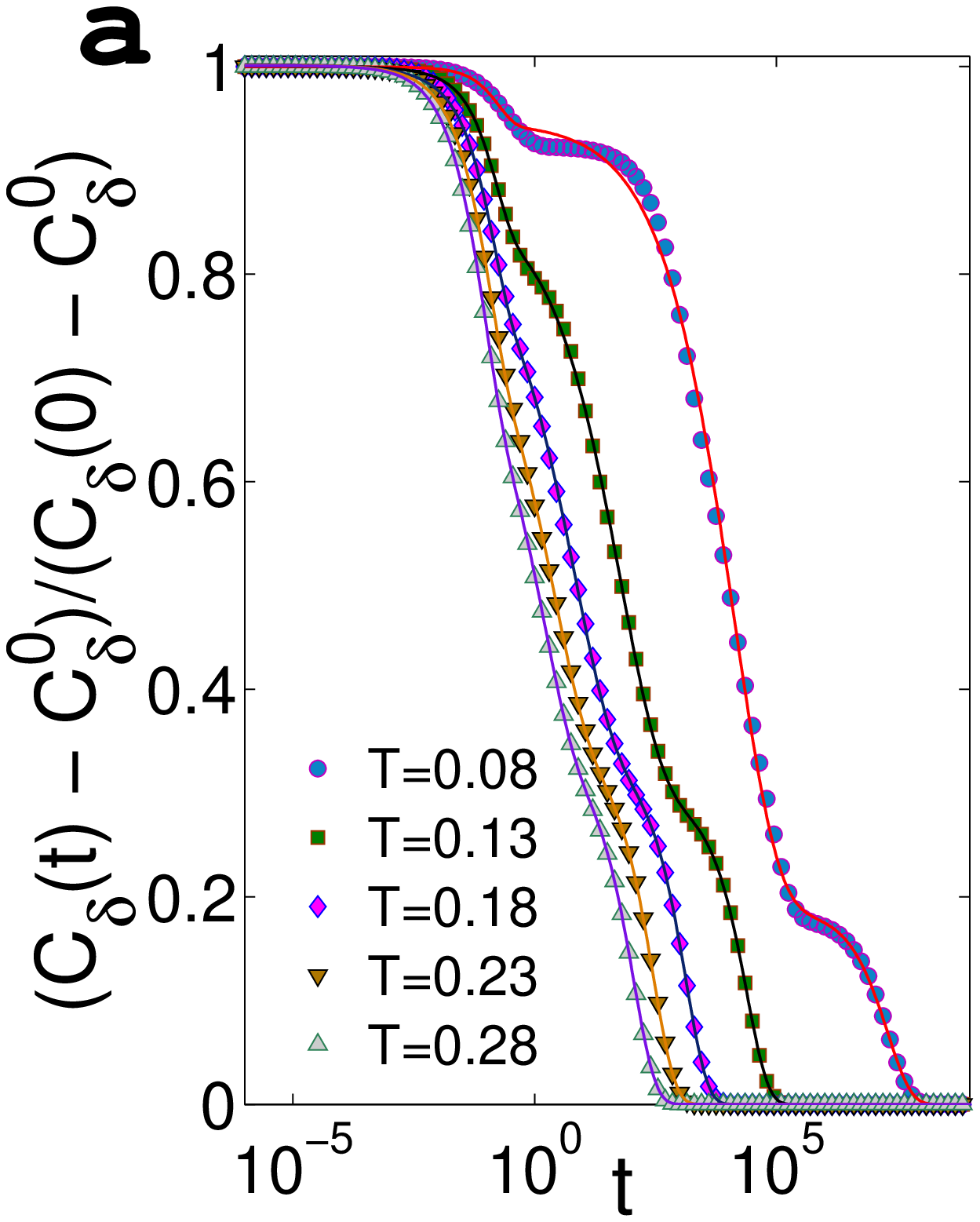} &
\includegraphics[width=4cm]{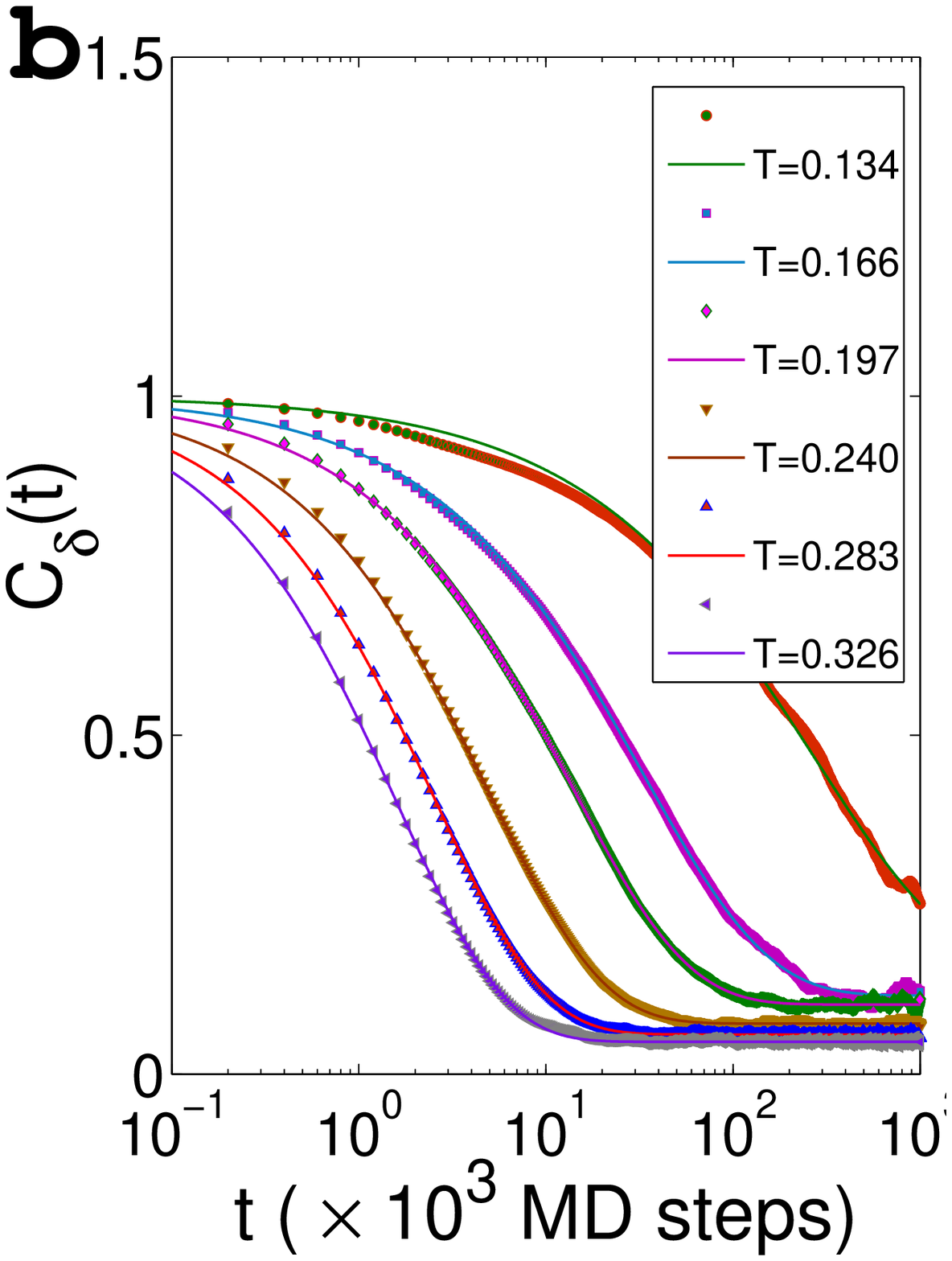}\\
\includegraphics[width=4cm]{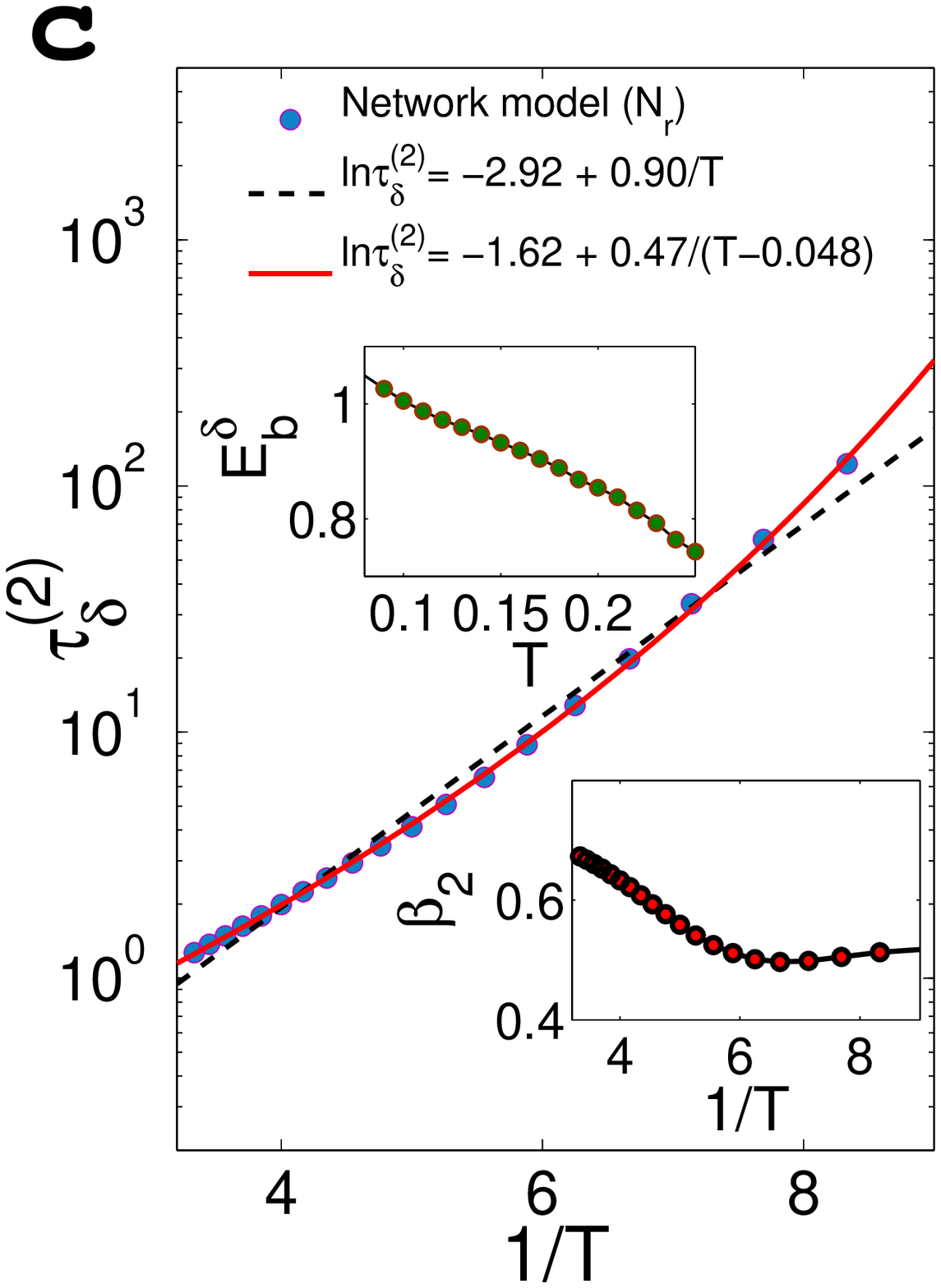}&
\includegraphics[width=4cm]{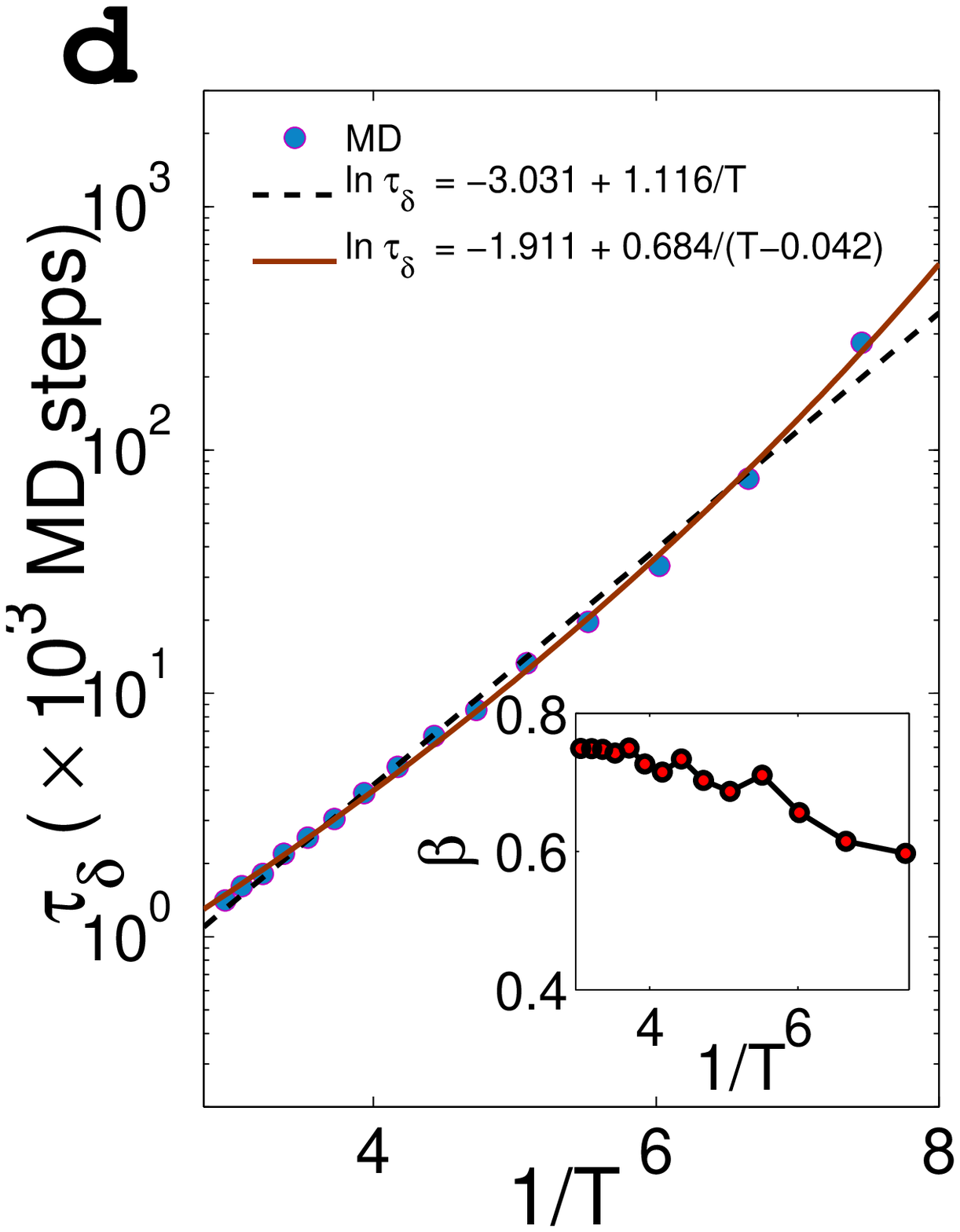}
\end{tabular}
\end{center}
\caption{Overlap function $C_\delta(t)$ for the network $\mathcal{N}_r$. Panel {\bf a}: $C_\delta(t)$ calculated in
the network model exhibits multi-step decay of correlation. Panel {\bf b}: $C_\delta(t)$ as obtained by confining
the MD trajectory in a part of the configuration space that excludes the basins of attraction of the first four
lowest-lying inherent structures. Panel {\bf c}: The intermediate relaxation time $\tau_\delta^{(2)}$ obtained by fitting 
the data for $C_\delta(t)$ in {\bf a} with the sum of three stretched exponentials. Here the $\ln\tau_\delta^{(2)}$
vs.~$1/T$ plot shows a small bending indicating deviations from the Arrhenius $T$-dependence. The VFT fit is also shown. the upper and
lower insets show, respectively,
the $E_b^\delta(T)$ calculated from Eq.\eqref{eq.EffectiveBarrier} and the exponent $\beta_2$ from the
fits to $C_\delta(t)$ at different temperatures. {\bf d}, The results for $\tau_\delta(T)$ obtained in MD simulations
agree with the \emph{fragile} behavior observed in the network model. The semiquantitative agreement can be verified by
comparing the VFT fits for $\tau_\delta(T)$ in the network model and in MD.}
\label{fig.OverlapCorrelation_Nr}
\end{figure}

\section{Comparison of the results of the network model with those of MD} \label{sec.MD}

We have carried out MD simulations \cite{DFrenkel} for the cluster of 13 particles interacting via the Morse potential
[Eq.\eqref{eq.MorsePotential}]. Initial long MD runs along with regular quenching provide the
starting data base of minima and then we follow the procedure of Appendix \ref{app.BuildNetwork} to improve the list 
as well as to
construct the network of minima and transition states. The main motivation of our MD study is to check to
what extent the dynamics of the systems is captured by the network model by comparing the results of the network
model calculation with those for the real dynamics (i.e.~Newton's equation of motion). We find that the main results of the network model calculation are supported both
qualitatively and quantitatively by the MD results. As already discussed in the introduction, the model assumes that there is a clear
separation between the time scales of local vibrations and activated jumps. This assumption is bound to be valid at very
low temperature where the barriers are much higher than the temperature. However, the barrier heights have a wide
distribution (Fig.\ref{fig.PESDetails} {\bf c}) and hence the assumption of well separated time scales may be invalidated
at different temperatures for different parts of the landscape. Presumably, as the temperature is increased,
the dynamics crosses over from a landscape dominated regime at low temperatures to a high-temperature 
regime where the system does not see the details of
the PEL due to its large thermal energy and the dynamics is no longer dominated by activation processes.

We have calculated the stress autocorrelation function $C_\sigma(t)$ and the overlap function
$C_\delta(t)$ (Sec.\ref{subsec.Structuralrelaxation}) from MD simulations at different temperatures. Also, we have 
estimated the
mean waiting times and waiting time distributions. We find that due to intra-basin relaxation present in MD,
the stress autocorrelation function $C_\sigma(t)$ decays within the vibrational time periods of the local
minima much before the inter-basin jump dynamics comes into play. As a result, $C_\sigma(t)$ is not useful for
comparison with the network model results. Hence we concentrate on $C_\delta(t)$ and the related relaxation
time since by definition the decay of $C_\delta(t)$ is entirely determined by inter-basin transitions. For
calculating this quantity as well as the waiting times for different minima, we construct a discontinuous
trajectory in terms of configurations of inherent structures from the MD trajectory and track the
transitions along it by the \emph{interval bisection} method described in reference \cite{BDoliwa1} [see Appendix
\ref{app.IntervalBisection}].

For comparing the results obtained for $\mathcal{N}_r$ (Sec.\ref{sec.NetworkMorseCluster}), additionally, we
need to confine the MD trajectories in a restricted part of the configuration space so that the basins of the
four lowest-lying minima are not visited (see Appendix \ref{app.IntervalBisection}). This goal is achieved by applying 
a procedure similar to that
described in Refs \cite{SFChekmarev1,SFChekmarev2}. The application of this method in conjunction with the interval bisection
method makes the MD runs very time consuming and hence it is difficult to get very long MD trajectories
(typically $10^8$ or $10^9$ MD steps with the MD step length $\delta t=0.001$). So we have averaged over many parallel runs with smaller
number of MD steps ($2\times 10^7$ for $\mathcal{N}_f$ and $2\times 10^6$ to $2\times 10^7$ for
$\mathcal{N}_r$, depending on the temperature) starting from different parts of the landscape i.e. different
initial conditions (different IS configurations are taken from the already existing data base). 

We compare the results for $C_\delta(t)$ obtained through the network model for $\mathcal{N}_f$ and
$\mathcal{N}_r$ with those obtained via \emph{isokinetic} MD \cite{DFrenkel} in Figs.\ref{fig.OverlapCorrelation_Nf} and
\ref{fig.OverlapCorrelation_Nr}, respectively. Fig.\ref{fig.OverlapCorrelation_Nf} {\bf a} shows the network
model results for $C_\delta$ in $\mathcal{N}_f$. Corresponding MD results are shown in 
Fig.\ref{fig.OverlapCorrelation_Nf} {\bf b}. The Arrhenius plots for associated relaxation times are shown in
Figs.\ref{fig.OverlapCorrelation_Nf} {\bf c} and {\bf d}. The \emph{strong} behavior is evident and decays of
$C_\delta(t)$ in both the cases are close to exponential ($\beta\simeq 0.9$) as exhibited in the insets of these
figure panels. A good estimate of the effective activation energy is once again obtained from
Eq.\eqref{eq.EffectiveBarrier} (inset of Fig.\ref{fig.OverlapCorrelation_Nf} {\bf c}). Here the effective
barrier $E_b^\delta\simeq 0.25$ is comparable to $T$ in the temperature range of interest and hence the associated 
time scale for relaxation is quite short. Even then, an accurate calculation of the correlation function
$C_\delta(t)$ and the relaxation time $\tau_\delta$ in MD is quite difficult because the system gets trapped in the
deep global minimum most of the time and a proper sampling of the relevant relaxation paths connecting the higher minima
to the global one becomes hugely time consuming and increasingly difficult with decreasing temperature. As a
result we can estimate $\tau_\delta$ only for a limited range of temperatures, as shown in
Fig.\ref{fig.OverlapCorrelation_Nf}{\bf d}. Also for temperatures higher
than $T\simeq E_b^\delta\simeq 0.25$ the network model dealing only with the activated processes becomes unreliable,
rendering a comparison with MD results inappropriate.

For the network $\mathcal{N}_r$, $C_\delta(t)$ exhibits a three-stage relaxation profile
(Fig.\ref{fig.OverlapCorrelation_Nr} {\bf a}) owing to the presence of well-separated time scales similar to the case 
of $C_\sigma(t)$ (Fig.\ref{fig.StressAutoCorrelation_Nr} {\bf b}). Such a clear cut separation of timescale is not
observed in the corresponding \emph{microcanonical} MD \cite{DFrenkel} results for $C_\delta(t)$ 
shown in Fig.\ref{fig.OverlapCorrelation_Nr} {\bf b}, although a
faint signature of multi-step relaxation can be seen at low temperatures. The decays of $C_\delta(t)$ at various
temperatures, shown in Fig.\ref{fig.OverlapCorrelation_Nr} {\bf a}, are fitted to a sum of three stretched exponentials 
[Eq.\eqref{eq.KWW}]. The temperature dependence of the intermediate
relaxation time $\tau_\delta^{(2)}(T)$ shows deviations from the Arrhenius form [Eq.\eqref{eq.Arrhenius}]
in Fig.\ref{fig.OverlapCorrelation_Nr} {\bf c}, although much less in extent than that for $\tau_\sigma^{(2)}(T)$ 
(Fig.\ref{fig.StressAutoCorrelation_Nr} {\bf c}). The latter fact is indicative of the importance of the quantity
$\Phi_{ab}$ of Eq.\eqref{eq.CorrelationFunction}, i.e.~the quantity whose autocorrelation function is used
to calculate the relaxation time, in determining the degree of fragility obtained from the temperature dependence of the
relaxation time. As discussed above, the distribution and temperature dependence of the quantities $\mathcal{W}_{ab}^\phi$, 
which depend on the quantity $\phi({\bf r})$ whose autocorrelation function is used
to calculate the relaxation time (see Appendix \ref{app.ActivationEnergy}), play an important role in determining the
degree of fragility. 
We have checked that plots of $\mathcal{W}_{ab}$ for the overlap function, similar to those shown in Fig. \ref{fig.MinimaPairWeight}, 
exhibit a less pronounced dependence on the temperature. This is consistent with the observation that the temperature dependence of 
the relaxation time extracted 
from the decay of the overlap function exhibits smaller deviations from the Arrhenius form (lower degree of fragility)
in comparison to that for the
relaxation time obtained from the stress autocorrelation function.
The Arrhenius fit parameter $E_b^\delta$ and VFT fit parameters
$B_\delta$ and $T_0^\delta$ are shown in Fig.\ref{fig.OverlapCorrelation_Nr} {\bf c}. These agrees reasonably well with 
the results of similar fits done for $\tau_\delta$ estimated through MD simulations
(Fig.\ref{fig.OverlapCorrelation_Nr} {\bf d}). 
 
\section{Conclusion}

Our work shows that the master equation based approach, first proposed by Angelani {\it et al.}\cite{LAngelani1}, for the 
landscape dominated activated dynamics on a network of minima and transition states is capable
of addressing many important issues and challenges related to glassy dynamics. Our results for the full network $\mathcal{N}_f$ are
similar to those of Refs.\cite{LAngelani1,LAngelani2} where {\emph strong} behavior in the dynamics, arising from the dominance 
of the global minimum and
the barriers surrounding it, was observed. Our study of the restricted network $\mathcal{N}_r$ leads to the important result
that the master equation approach may lead to fragile dynamic behavior if many inherent structures and the barriers between pairs of them
are involved in the relaxation process. This conclusion is confirmed by our MD simulations. 
Our study, thus, provides valuable insights into the origin of fragile dynamic behavior in glass-forming liquids.
It also illustrates the usefulness of the master equation approach in studies of glassy dynamics. Due to computational constraints,
the applicability of this approach has been limited to small system sizes till now. A related
approach \cite{AKushima} that differs from the one considered here in the details of its implementation seems to be a 
promising one for studying larger systems.
Another interesting future direction for studying moderately large systems might be to extend this framework for the
dynamics in the \emph{metabasin} space to obtain quantities relevant for the characterization
of glassy dynamics.      

\textbf{Acknowledgements} 

We thank Smarajit Karmakar, Pinaki Chaudhury and Biswaroop Mukherjee for useful discussions. SB would like to
acknowledge CSIR (Government of India) and DST (Govt.~of India) for support. CD acknowledges support from DST(Govt. of India).

\appendix
\section{Effective activation energy}\label{app.ActivationEnergy}
 As mentioned in Section \ref{subsec.CorrelationFunction}, $\tau_n=|\lambda_n|^{-1}$ is the relaxation time
corresponding to the $n$-th mode ($n\geq 2$) and $w_\phi^{(n)}$ [Eq.\eqref{eq.ModeWeight}] provides the distribution of such
relaxation times. We assume that the relaxation time for each individual mode is governed by an effective barrier
$E_b^{(n)}$ i.e.~$\tau_n\propto \exp{(E_b^{(n)}/T)}$. Hence we can estimate $E_b^{(n)}$ approximately from
the local slope of the $\ln\tau_n$ vs.~$1/T$ curve as follows
\begin{subequations}
\begin{eqnarray}
&&E_b^{(n)}\simeq-\frac{1}{\lambda_n}\frac{\partial |\lambda_n|}{\partial(1/T)} \nonumber\\
&&\simeq\frac{1}{|\lambda_n|}\sum_{<ab>,s}\left[(V_{ab}^s-V_a)(\mathcal{U}_{s,a}^{(n)})^2+(V_{ab}^s-V_b)(\mathcal{U}_{s,b}^{(n)})^2\right.\nonumber\\
&&\left.-2(V_{ab}^s-\frac{V_a+V_b}{2})\mathcal{U}_{s,a}^{(n)}\mathcal{U}_{s,b}^{(n)}\right]
\label{eq.ModeActivationEnergy}
\end{eqnarray}
with
\begin{eqnarray}
\mathcal{U}_{s,a}^{(n)}&=&(\Lambda_{ab}^s)^{1/2}\left[\frac{\mathrm{Det}(\mathbf{H}_a)}{\mathrm{Det}(\mathbf{H}_{ab}^s)}\right]^{1/4}e^{-\frac{V_{ab}^s-V_a}{2T}}
e_a^{(n)}
\end{eqnarray} 
\end{subequations}
To obtain the expression \eqref{eq.ModeActivationEnergy} for $E_b^{(n)}$ we have rewritten $\lambda_n$ as
$\sum_{a,b}\widetilde{W}_{ab}e_a^{(n)}e_b^{(n)}$ and while taking the derivative of $\lambda_n$ with respect
to $T$ neglected the temperature dependence of the eigenvector
$\mathbf{e}^{(n)}$ assuming it to be weakly temperature dependent. We find this to be valid for the case of
the 13-atom Morse cluster and the above approximation for $E_b^{(n)}$ seems to work, as we have reported in
Sections \ref{sec.DynamicsMorseCluster} and \ref{sec.MD}.
The overall effective activation energy $E_b^\phi(T)$ is obtained by summing over the contributions of all the
modes appearing with the weights $w_\phi^{(n)}$ [Eq.\eqref{eq.ModeWeight}], i.e.~$E_b^\phi\equiv\sum_{n\geq
2}E_b^{(n)}w_\phi^{(n)}$. This can be recast as
\begin{subequations}
\begin{eqnarray}
E_b^\phi&=&\sum_{<ab>}\mathcal{W}_{<ab>}^\phi E_{<ab>}^\phi \label{eq.EffectiveBarrier}
\end{eqnarray}
here,
\begin{eqnarray}
\mathcal{W}_{<ab>}^\phi&\equiv&\sum_s \mathcal{W}_{s,<ab>}^\phi \label{eq.MinimaPairWeight}\\
E_{<ab>}^\phi&\equiv&\frac{\sum_s \mathcal{E}_{s,<ab>}^\phi}{\sum_s \mathcal{W}_{s,<ab>}^\phi}
\end{eqnarray}
with
\begin{eqnarray}
&&\mathcal{W}_{s,<ab>}^\phi=\sum_{n\geq
2}|\lambda_n|^{-1}w_\phi^{(n)}(\mathcal{U}_{s,a}^{(n)}-\mathcal{U}_{s,b}^{(n)})^2 \nonumber\\
&&\mathcal{E}_{s,<ab>}=\sum_{n\geq
2}|\lambda_n|^{-1}w_\phi^{(n)}\left[V_{ab}^s(\mathcal{U}_{s,a}^{(n)}-\mathcal{U}_{s,b}^{(n)})^2\right.
\nonumber \\
&&\left.-(V_a\mathcal{U}_{s,a}^{(n)}-V_b\mathcal{U}_{s,b}^{(n)})(\mathcal{U}_{s,a}^{(n)}-\mathcal{U}_{s,b}^{(n)})\right]\nonumber
\end{eqnarray}
\end{subequations}
The effective barrier $E_b^\phi(T)$ obtained from Eq.\eqref{eq.EffectiveBarrier} compares quite well with the
barrier extracted from the $T$-dependence of the relaxation time $\tau_\phi$ estimated by fitting the KWW form
[Eq.\eqref{eq.KWW}] to the correlation function $C_\phi(t)$, computed using Eq.\eqref{eq.CorrelationFunction}
for the Morse cluster. The quantity $\mathcal{W}_{ab}^\phi=\mathcal{W}_{ba}^\phi\equiv\mathcal{W}_{<ab>}^\phi/2$
can be a good measure of the relative importance of a pair of minima or an \emph{elementary jump} in the
relaxation process as demonstrated in Section \ref{sec.DynamicsMorseCluster}. 
 
\section{Construction of the network of minima and transition states} \label{app.BuildNetwork}

As mentioned in Sec.\ref{sec.NetworkMorseCluster}, we follow the procedure similar to that in
Ref.~\cite{MAMiller2} for building the network model for the 13-atom Morse cluster. 
We have used the OPTIM package \cite{Optim}, developed by D. J. Wales and co-workers, to search for
minima and transition states as follows
\begin{enumerate}
\item Do long MD simulations and steepest descent quenches in regular intervals to reach nearby minima
along the MD trajectories. Distinction between configurations of minima is done in terms of their potential
energy values. This provides us with an initial data base or list of minima.
\item Starting from each of these minima, one by one,
\begin{enumerate}
\item search for a transition state along the eigenvector with the lowest eigenvalue.
\item After reaching a transition state, do steepest descent minimization starting parallel to the transition 
vector, i.e.~eigenvector corresponding to the negative eigenvalue, at the transition state
to arrive at the minima connected directly to initial one (sometimes, following the eigenvector from a minimum we
obtain states which are not connected to it by steepest descent - we discard these states). This
establishes one edge, constituted of a pair of minima connected by a transition state, of the network.
\item Repeat (a) beginning anti-parallel to the eigenvector with the lowest eigenvalue and then successively in
both directions along eigenvectors with ascending eigenvalues until all the eigendirections are considered at
the starting minimum.
\item If some new minima, not in the starting list, are found in this process we append them to the data base
and modify the existing list of minima.
\end{enumerate}
\item Repeat 2 until all the minima in the data base are searched.
\end{enumerate}

\section{Interval bisection and confinement of MD trajectories in a specified part of the PEL}
\label{app.IntervalBisection}

Starting from a MD trajectory, $\mathbf{r}(t)=(\mathbf{r}_1(t),\mathbf{r}_2(t),...,\mathbf{r}_N(t))$, we
construct a trajectory
$\mathbf{r}^0(t)=(\mathbf{r}_1^0(t),\mathbf{r}_2^0(t),...,\mathbf{r}_N^0(t))$ in terms of the
inherent structure configurations by doing steepest descent minimization at regular intervals \cite{BDoliwa1}. The
straightforward way, though computationally impractical, is to do minimization at every time step. Also there
may be many back and forth jumps between neighboring minima giving rise to events that do not affect the
long time relaxation which we are mainly interested in. Bearing this fact in mind, we do quenches for
equidistant points, $t_n=n\Delta t$ ($n$ being an integer) to get $\mathbf{r}^0(t_n)$ with $\Delta
t=100-200$ MD steps (around $1/10$-th of the typical vibrational period at a minimum for the Morse cluster). During a MD run, starting from initial point ($n=0$),
\begin{enumerate}
\item Save the trajectory $\mathbf{r}(t)$ between $t_i=t_n$ and $t_f=t_{n+1}$ and then follow the steps
described below provided $\mathbf{r}^0(t_i)\neq \mathbf{r}^0(t_f)$ (actually we compare
$V(\mathbf{r}^0)$ instead of $\mathbf{r}^0$),
\begin{enumerate}
\item Set $t_m=(t_i+t_f)/2$ and get $\mathbf{r}^0(t_m)$ from $\mathbf{r}(t_m)$.
\item If $\mathbf{r}^0(t_m)=\mathbf{r}^0(t_i)$, set $t_i=t_m$ else set
$t_f=t_m$.
\item Go on repeating steps (a) and (b) until $t_f-t_i=1$. 
\end{enumerate}
\item Change $n$ to $n+1$ and go back to 1.
\end{enumerate}

Next we describe the method that we have adopted for restricting the trajectory of the state point of the
system to a part of the configuration space. This is an extension of the interval bisection method described
in the preceding paragraph. The trajectory is constructed part by part successively and in each part, whenever we 
detect a transition (using the
interval bisection method) from the \emph{allowed} part to \emph{forbidden} part
(say the part containing the basins of attraction of the four lowest-lying minima with $V_a<-43$ in the case of
the 13-atom Morse cluster), we re-initiate the MD at the point when the system last visited the allowed part using new
velocities for the particles. We assign random velocities drawn from a Maxwell distribution at temperature
$T$ [in the case of microcanonical simulation at constant $E_\mathrm{tot}$, the temperature is obtained from the total kinetic
energy at the instant when the system last visited the allowed part i.e.~$T=2(E_\mathrm{tot}-V)/(3N-6)$]. After each
such re-initiation we correct for the integrals of motion i.e.~energy, momentum and angular momentum by
suitable rescaling and rotation of the velocities. 

Under this procedure the dynamics no longer remains truly Newtonian due to repeated velocity re-initiations. 
We have calculated the velocity autocorrelation function which decays very fast (within 2000 MD steps
i.e.~$\sim$1-2 
vibrational periods) and heuristically we can argue that successive velocity re-initiations are more or less
uncorrelated. Nonetheless, following reference \cite{SFChekmarev2} we have performed a few standard checks by computing quantities that
can be accessed through the above mentioned \emph{confined} MD as well as through the regular unrestricted MD
procedure and we found good agreement between the results obtained in these two different ways. For instance, the distributions 
of the total kinetic energy and the waiting time for a restricted part of the
configuration space (e.g.~the basin of attraction of a particular minimum) can be estimated \cite{SFChekmarev2} both by enforcing the
trajectory to sample only the specified part or by picking out from a conventional long MD trajectory the
intervals during which the system samples the specified part. 

We find that there are some spurious effects in the
dynamics due to cycling of trajectories near the saddles or the system spending more time near the border of
the allowed and forbidden regions. Also, at low temperatures, when the system is constrained in the high
energy parts such as $\mathcal{N}_r$, it has a tendency to visit the lowest lying basins more often. In
the above mentioned procedure the system can visit the forbidden region within the small interval $\Delta t$ and come
back to the allowed part. We find these events to affect the properties related to short time relaxation such
as the diffusion constant [Eq.\eqref{eq.DiffusionConstant}], waiting time $\tau_w$ [Eq.\eqref{eq.WaitingTime}] and its 
distribution. While restricting the system in $\mathcal{N}_r$, as the system frequently escapes to the four
lower lying basins, the barrier that appears in the
temperature dependence of $D$ and $\tau_w$ turns out to be related to the barriers connecting $\mathcal{N}_r$
to the forbidden low-energy part. However, since these visits are really short lived ($\leq 100-200$ MD
steps), they do not affect
the features of long-time relaxation, such as the decay of correlations characterized by $C_\delta(t)$.


\begin{thebibliography}{10}

\bibitem{CAAngell1} C.~A.~Angell, J.~Phys.: Condens.~Matter {\bf 12}, 6463 (2000).
\bibitem{WKob} See the article by W.~Kob in \emph{Slow Relaxations and Nonequilibrium Dynamics in Condensed
Matter}, edited by J.-L.~Barrat, M.~Feigelman, and J.~Dalibard (Springer, Berlin, 2003).
\bibitem{PGDebenedetti} P.~G.~Debenedetti and F.~H.~Stillinger, Nature {\bf 419}, 259 (2001).
 \bibitem{CAAngell2} C.~A.~Angell, P.~H.~Poole and J.~Shao, Nuovo Cimento D {\bf 16}, 993 (1994).
\bibitem{GRuocco1} G.~Ruocco, F.~Sciortino, F.~Zamponi, C.~De Michele and T.~Scopigno, J.~Chem.~Phys.~{\bf
120}, 10666 (2004).
\bibitem{MGoldstein1} M.~Goldstein, J.~Chem.~Phys.~{\bf 51}, 3728 (1969).
\bibitem{DJWales1} D.~J.~Wales, \emph{Energy Landscapes} (Cambridge University Press, Cambridge, 2003).
\bibitem{KBinder} K.~Binder and A.~P.~Young, Rev.~Mod.~Phys.~{\bf 58}, 801 (1986).
\bibitem{FHStillinger1} F.~H.~Stillinger and T.~A.~Weber, Phys.~Rev.~A {\bf 25}, 978 (1982). 
\bibitem{FHStillinger2} F.~H.~Stillinger and T.~A.~Weber, Phys.~Rev.~A {\bf 28}, 2408 (1983).
\bibitem{FHStillinger3} F.~H.~Stillinger, Science {\bf 267}, 1935 (1995).
\bibitem{FHStillinger4} F.~H.~Stillinger, J.~Chem.~Phys.~{\bf 88}, 7818 (1988).
\bibitem{SSastry1} S.~Sastry, P.~G.~Debenedetti and F.~H.~Stillinger, Nature {\bf 393}, 554 (1998).
\bibitem{DFrenkel} D.~Frenkel and B.~Smit, \emph{Understanding Molecular Simulation} (Academic Press, San
Diego, 1996).
\bibitem{LAngelani1} L.~Angelani, G.~Parisi, G.~Ruocco and G.~Viliani, Phys.~Rev.~Lett.~{\bf 81}, 4648 (1998).
\bibitem{LAngelani2} L.~Angelani, G.~Parisi, G.~Ruocco and G.~Viliani, Phys.~Rev.~E {\bf 61}, 1681 (2000).
\bibitem{MAMiller1} M.~A.~Miller, J.~P.~K.~Doye and D.~J.~Wales, Phys.~Rev.~E {\bf 60}, 3701 (1999).
\bibitem{FHStillinger5} F.~H.~Stillinger, Phys.~Rev.~E {\bf 59}, 48 (1999).
\bibitem{BDoliwa1} B.~Doliwa and A.~Heuer, Phys.~Rev.~E {\bf 67}, 031506 (2003).
\bibitem{AHeuer} A.~Heuer, J.~Phys.:Condens. Matter {\bf 20}, 373101 (2008).
\bibitem{YYang} Y.~Yang and B.~Chakraborty, Phys.~Rev.~E {\bf 80}, 011501 (2009).
\bibitem{TFMiddleton} T.~F.~Middleton and D.~J.~Wales, Phys.~Rev.~B {\bf 64}, 024205 (1999).
\bibitem{JPKDoye1} J.~P.~K.~Doye and C.~P.~Massen, J.~Chem.~Phys.~{\bf 122}, 084105 (2005).
\bibitem{PMMorse} P.~M.~Morse, Phys.~Rev.~{\bf 34}, 57 (1929).
\bibitem{MAMiller2} M.~A.~Miller, J.~P.~K.~Doye and D.~J.~Wales, J.~Chem.~Phys.~{\bf 110}, 328 (1999).
\bibitem{SFChekmarev1} S.~F.~Chekmarev, Phys.~Rev.~E {\bf 64}, 036703 (2001).
\bibitem{SFChekmarev2} S.~F.~Chekmarev and S.~V.~Krivov, Chem.~Phys.~Lett.~{\bf 287}, 719 (1998).
\bibitem{WHPress} W.~H.~Press, S.~A.~Teukolsky, W.~T.~Vetterling and B.~P.~Flannery, \emph{Numerical Recipes
in C} (Cambridge University Press, Cambridge, 1992).
\bibitem{CJCerjan} C.~J.~Cerjan and W.~H.~Miller, J.~Chem.~Phys.~{\bf 75}, 2800 (1981).
\bibitem{Optim} http://www-wales.ch.cam.ac.uk/software.html.
\bibitem{HAKramers} H.~A.~Kramers, Physica {\bf 7}, 284 (1940).
\bibitem{JSLanger} J.~S.~Langer, Ann.~Phys.~{\bf 54}, 258 (1969).
\bibitem{PHanggi} P.~H$\mathrm{\ddot{a}}$nggi, P.~Talkner and M.~Borkovec, Rev.~Mod.~Phys.~{\bf 62}, 251 (1990).
\bibitem{RKohlrausch} R.~Kohlrausch, Ann.~Phys.~(Leipzig) {\bf 12}, 393 (1847).
\bibitem{GWilliams} G.~Williams and D.~C.~Watts, Trans.~Faraday Soc.~{\bf 66}, 80 (1980).
\bibitem{JPHansen} J.~P.~Hansen and I.~R.~McDonald, \emph{Theory of Simple Liquids} (Academic Press, San
Diego, 1976).
\bibitem{BDoliwa2} B.~Doliwa and A.~Heuer, Phys.~Rev.~E {\bf 67}, 030501 (2003).
\bibitem{HVogel} H.~Vogel, Phys.~Z.~{\bf 22}, 645 (1921)
\bibitem{GSFulcher} G.~S.~Fulcher, J.~Amer.~Ceram.~Soc.~{\bf 8}, 339 (1925).
\bibitem{GTammann} G.~Tammann and W.~Z.~Hesse, Anorg.~Allg.~Chem.~{\bf 156}, 245 (1926).
\bibitem{AKushima} A.~Kushima et al., J.~Chem. Phys.~{\bf 130}, 224504 (2009).
\end{thebibliography}
\end{document}